\documentclass{article}

\PassOptionsToPackage{numbers, compress}{natbib}
\usepackage{orcidlink}
\usepackage{graphicx} 
\usepackage{amsmath}
\usepackage{booktabs}
\usepackage{float}

\usepackage{algorithm}
\usepackage{algpseudocode}
\usepackage{amsfonts}

\usepackage[preprint]{neurips_2024}

\usepackage[utf8]{inputenc} % allow utf-8 input
\usepackage[T1]{fontenc}    % use 8-bit T1 fonts
\usepackage{hyperref}       % hyperlinks
\hypersetup{
    pdfborder={0 0 0},        % Remove the border around links
    colorlinks=true,          % Enable colored links
    linkcolor=blue,           % Link color for internal document links
    urlcolor=blue,            % Link color for external URLs
    citecolor=blue            % Link color for citations
}
\usepackage{url}            % simple URL typesetting
\usepackage{booktabs}       % professional-quality tables
\usepackage{amsfonts}       % blackboard math symbols
\usepackage{nicefrac}       % compact symbols for 1/2, etc.
\usepackage{microtype}      % microtypography
\usepackage{xcolor}         % colors
\usepackage{xr}             % external document referencing

\title{dZiner: Rational Inverse Design of\\ Materials with AI Agents}

\author{%
  \orcidlink{0000-0001-5696-9193} Mehrad Ansari\thanks{Denotes corresponding author.} \\
  Acceleration Consortium\\
 University of Toronto\\80 St George St, Toronto, ON M5S 3H6 \\
  \texttt{mehrad.ansari@utoronto.ca} \\
  \And
  \orcidlink{0000-0002-5352-0419} Jeffrey Watchorn \\
  Acceleration Consortium\\
  University of Toronto\\
  80 St George St, Toronto, ON M5S 3H6\\
  \texttt{jeff.watchorn@utoronto.ca}
  \And
    \orcidlink{0009-0007-7583-4148} Carla E. Brown \\
  Acceleration Consortium\\
  University of Toronto\\
  80 St George St, Toronto, ON M5S 3H6\\
  \texttt{carla.brown@utoronto.ca}
  \And
  \orcidlink{0000-0002-8145-800X} Joseph S. Brown \\
  Acceleration Consortium\\
  University of Toronto\\
  80 St George St, Toronto, ON M5S 3H6\\
  \texttt{js.brown@utoronto.ca}
}

\begin{document}
\maketitle
\begin{abstract}
  Recent breakthroughs in machine learning and artificial intelligence, fueled by scientific data, are revolutionizing the discovery of new materials.
  Despite the wealth of existing scientific literature, the availability of both structured experimental data and chemical domain knowledge that can be easily integrated into data-driven workflows is limited. 
The motivation to integrate this information, as well as additional context from first-principle calculations and physics-informed deep learning surrogate models, is to enable efficient exploration of the relevant chemical space and to predict structure-property relationships of new materials \textit{a priori}. 
Ultimately, such a framework could replicate the expertise of human subject-matter experts.
In this work, we present dZiner, a chemist AI agent, powered by large language models (LLMs), that discovers new compounds with desired properties via inverse design (property-to-structure). 
In specific, the agent leverages domain-specific insights from foundational scientific literature to propose new materials with enhanced chemical properties, iteratively evaluating them using relevant surrogate models in a rational design process, while accounting for design constraints.
The model supports both closed-loop and human-in-the-loop feedback cycles enabling human-AI collaboration in molecular design with real-time property inference, and uncertainty and chemical feasibility assessment.
We demonstrate the flexibility of this agent by applying it to various materials target properties including surfactants, ligand and drug candidates, and metal-organic frameworks. 
Our approach holds promise to both accelerate the discovery of new materials and enable the targeted design of materials with desired functionalities.
The methodology is available as an open-source software on \href{https://github.com/mehradans92/dZiner}{\color{blue}{{https://github.com/mehradans92/dZiner}}}.

\end{abstract}
\section{Introduction}
The discovery of new molecules and materials with advanced properties is essential for tackling significant challenges, ranging from therapeutic discovery to addressing climate change.
The evolution of materials innovation has gone through four distinct paradigms~\cite{agrawal2016perspective}.
Initially, it primarily relied on empirical trial and error.
And then, as disciplines like mathematics, chemistry, and physics advanced, materials innovation began to follow scientific laws.
The third paradigm emerged with the advent of computational chemistry, illustrated by tools such as Gaussian 70 for ab initio calculations and density functional theory (DFT)~\cite{pople1999quantum, garrity2014pseudopotentials}.
Currently, the fourth paradigm integrates theoretical, experimental, and computational methodologies using data-driven techniques including data mining, cluster analysis, predictive analytics, machine learning (ML), and materials informatics altogether~\cite{mosavi2012reactive, samudrala2013informatics}.

One major drawback of the traditional materials discovery methods is that it often involves extensive screening through laboratory experiments or in silico simulations, which are both time-consuming and resource-intensive~\cite{hautier2012computer, pyzer2015high}.
On the other hand, the promising data-driven approaches that use machine learning surrogate models to predict material structures and properties~\cite{merchant2023scaling, chen2022universal, ansari2024learning} or suggest novel materials~\cite{ren2022invertible, zeni2023mattergen} rely heavily on extensive training datasets. 
However, these models face challenges when such data is unavailable or when there is only a limited budget for conducting experiments or simulations.
In contrast, a human expert would be much more effective in such cases, by leveraging their domain knowledge and reasoning from limited examples.
This underscores the need for a new materials design paradigm, where models are built to replicate and/or integrate the expertise of human domain experts.

The emergence of large language models (LLMs), which excel at understanding and processing natural language text, presents a promising opportunity to integrate primary sources from complex scientific literature, diverse datasets, and human expertise toward the acceleration of scientific discoveries.
LLMs have excelled at various tasks, even those they are not explicitly trained for, which has led to increasing interest in creating LLM-based agents with abilities including human-mimicking reasoning, self-reflection, and decision-making~\cite{wei2022chain, huang2022towards, li2022pre}.
These autonomous agents can be augmented with external tools or action modules, enabling them to surpass conventional text processing and directly interact with the physical world (i.e. robotic manipulation~\cite{huang2023voxposer, brohan2023can, darvish2024organa} and scientific experimentation~\cite{boiko2023autonomous, m2024augmenting}). 
By integrating tools such as plugins specific to domain expertise, these agents can overcome the inherent deficiencies of LLMs in specific domains and enhance their overall applicability, performance, and interpretability~\cite{wellawatte2023extracting, kim2024explainable}.
For instance, recent studies have demonstrated the use of LLM agents to extract materials datasets and scientific research~\cite{ansari2023agent,lala2023paperqa, schilling2024text, leong2024automated, skarlinski2024languageagentsachievesuperhuman}, chemical innovation~\cite{chen2023towards},
experiment planning~\cite{prince2023opportunities} and predicting experimental outcomes~\cite{ramos2023bayesian}, hypothesis generation~\cite{gao2024empowering, gu2024generation}, and closed-loop or human-in-the-loop molecular discovery~\cite{jia2024llmatdesign, mcnaughton2024cactus}, among many other applications.
An excellent overview of LLM-based autonomous agents in chemistry and materials can be found in reference~\cite{ramos2024review}.

\begin{figure}[t!]
    \centering
    % \captionsetup{justification=centering}
    \includegraphics[width=\textwidth]{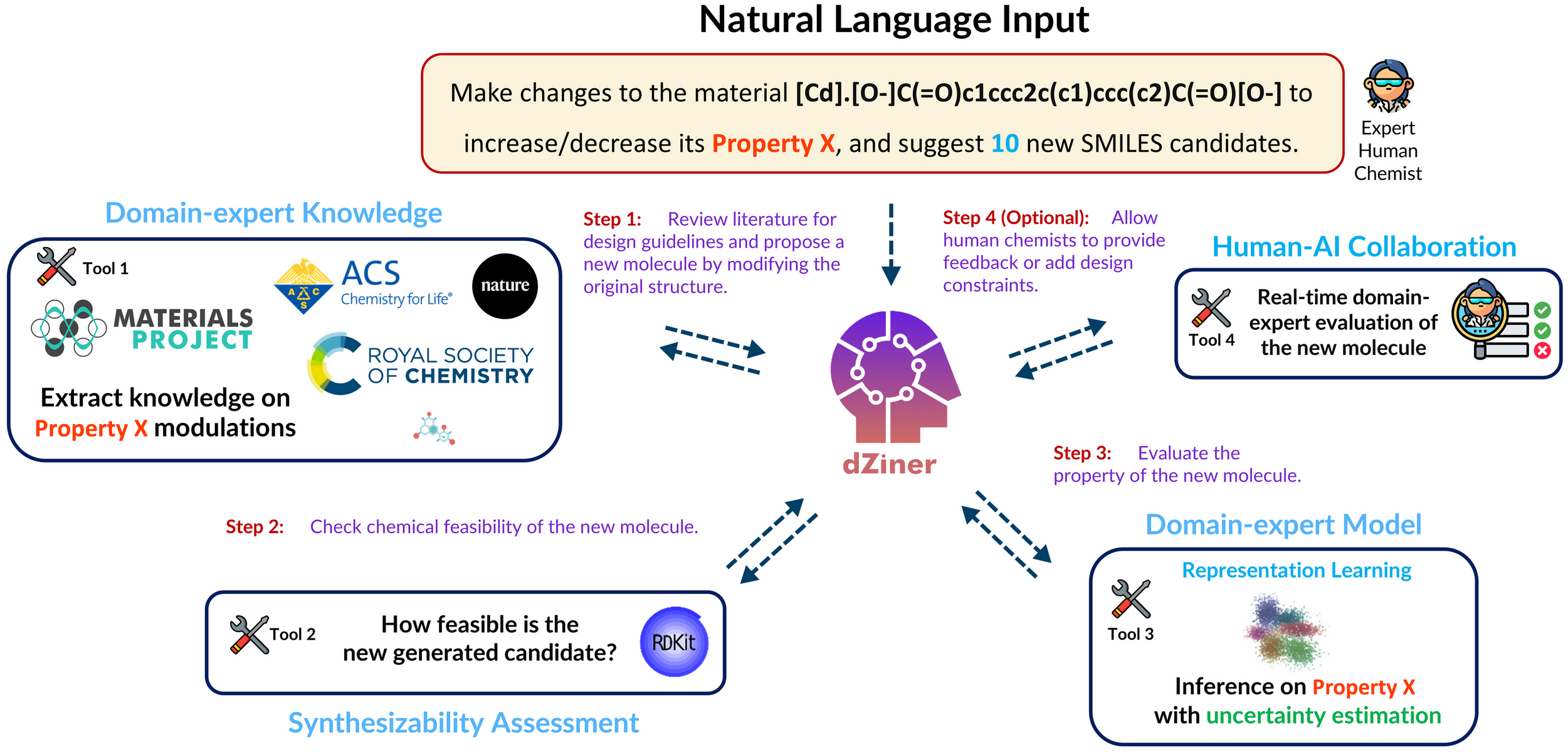}
    \caption{\textbf{dZiner workflow overview}.
    The model starts by inputting the material's initial structure as a textual representation. 
    The AI agent dynamically retrieves domain-knowledge (design guidelines) for Property X from scientific literature, the Internet or other resources. 
    Based on these guidelines, and any additional design constraints provided in \emph{natural language}, the agent proposes a new candidate and assesses its chemical feasibility in real-time.
    Next, it estimates Property X for the new candidate, incorporating epistemic uncertainty, using a cost-efficient surrogate model. 
    Optionally, as part of a human-in-the-loop process, the human chemist can review the agent's new candidates and chain-of-thoughts, providing feedback and suggesting further modifications or constraints, creating an opportunity for human-AI collaboration to guide the exploration process.
    The agent continues exploring the chemical space, guided by chemistry-informed rules, until it meets the convergence criteria.
 \label{fig:toc}}
\end{figure}

In this work, we present dZiner, an agent-based framework for rational inverse design of materials, powered by the state-of-the-art LLMs (Figure~\ref{fig:toc}).
Leveraging both human expertise and the existing knowledge contained in scientific publications, dZiner acts as an intelligent chemist research assistant~\cite{mirza2024large}, providing feedback on every step of the iterative inverse design process. 
Our agent starts by inputting the initial molecule as a textual representation, SMILES (Simplified Molecular Input Line-Entry System) or a sequence string, along with a brief description of the property optimization task (e.g., increase binding affinity, decrease critical micelle concentration (CMC), or increase CO$_2$ adsorption), all in natural language.
The agent then interprets human-provided instructions or any design constraints, along with retrieving chemical knowledge from relevant scientific literature, the Internet or other resources to identify possible chemical  modifications that could potentially improve the target molecular property. %to the functional groups or elements within the molecule that could lead to a favorable change, based on the desired property optimization task.
Following these modulation guidelines, the agent generates a new candidate molecule.
However, since SMILES strings generated by LLMs may sometimes deviate from proper SMILES grammar, resulting in invalid structures or potential hallucinations, we implement a validation step. 
This step serves as a quick check to assess the chemical feasibility, and score synthesizability of the newly generated molecule.
After this validation process, the effectiveness of the molecule modulation is assessed using a domain-expert model, potentially physics-based.
However, %given the high cost of nature of such models (e.g. DFT or Free Energy Perturbation calculations), 
to reduce the computational cost of the framework, we limit our study to use more affordable surrogate data-driven models for evaluation rather than expensive DFT or Free Energy Perturbation calculations.
The agent then iteratively reviews the modified materials and the entire modification history, stopping the generation of new candidates once the convergence criteria are met.
Optionally, in a human-in-the-loop process, a chemist can review the agent's proposed candidates and reasoning, offering feedback and suggesting additional modifications or constraints. 
This enables human-AI collaboration, allowing the chemist to better guide and refine the exploration process.

This manuscript is structured as follows: Section~\ref{sec:results} presents the benchmarking and evaluation of the model's performance across three distinct materials inverse design tasks. 
Section~\ref{sec:discission} follows with a discussion on the implications of our findings, the strengths and limitations of our approach, and potential directions for future research. 
Finally, Section~\ref{sec:methods} provides details of our methodology, including the agent's toolkits, domain-expert knowledge and surrogate models, and synthesizability assessment.
% Visualization for the 600 AI-generated molecules, are available in the Supplementary Information.
\section{Results}
\label{sec:results}
\subsection{Surfactant Design and Critical Micelle Concentration Inference\label{sec:results_CMC}}
Surfactant molecules play important roles in a wide variety of disciplines of study, from lubricants and coating to pharmaceuticals and drug delivery systems \cite{de2015review}. 
This wide applicability of study is due to the role of surfactant molecules which act as compatibilizers between dissimilar materials phases. 
While there are many metrics that are used to characterize surfactant molecules, the most common is the critical micelle concentration (CMC). 
CMC is traditionally the experimentally determined concentration at which individual surfactant molecules will self-assemble into larger aggregates (micelles). 
This value is critically important as the desirable properties of surfactants (solubilizing differing phases, enabling biocompatibility etc.) are typically only enabled when the solution concentration of the surfactant is above the CMC~\cite{perinelli2020cmc}. 
To design surfactant molecules with a desired CMC, the task is often challenging and relies heavily on domain-knowledge based expertise. Hence, the design task of minimizing CMC is both well-suited for an LLM agent, and a desirable objective to reduce the reliance on domain expertise for chemical synthesis.

Given these considerations, we apply dZiner to the rational design of surfactant molecules, with the objective of generating synthesizable molecules that minimize their expected CMC in water at room temperature. 
The agent was provided with an initial candidate surfactant-like molecule, for these experiments N-(2-oxotetrahydrofuran-3-yl) decanamide, and was tasked with making additions, substitutions or deletions to reduce CMC. 
The expected CMC with uncertainty is evaluated via a surrogate model as outlined in the methods  (section~\ref{methods:cmc}).
The design guidelines were determined by the agent via providing exemplary journal articles \cite{czajka2015surfactants,gaudin2016new,mozrzymas2011prediction,huibers1997prediction,li2004estimation,xuefeng2006correlation,moriarty2023analyzing,boukelkal2024qspr} on surfactant design.
These general guidelines include; 1. hydrophobic tail length and structure; increasing the length of the tail generally reduced CMC while increasing branching reduces CMC, 2. hydrophilic head group size and polarity; larger and more polar head groups generally increase CMC by increasing aqueous solubility, 3. functionalization with heteroatoms, aromatic moieties or other functional groups; modifications to add silicons, fluorines or other groups such as ethylene oxides to the tail or head respectively, reduces CMC. Additionally, the model is asked to keep the molecular weight of the generated candidates lower than 600 (g/mol) in natural language text.

With this information, the agent was applied to the inverse design task. 
\begin{figure}[b!]
    \centering
    % \captionsetup{justification=centering}
        \includegraphics[width=0.95\textwidth]{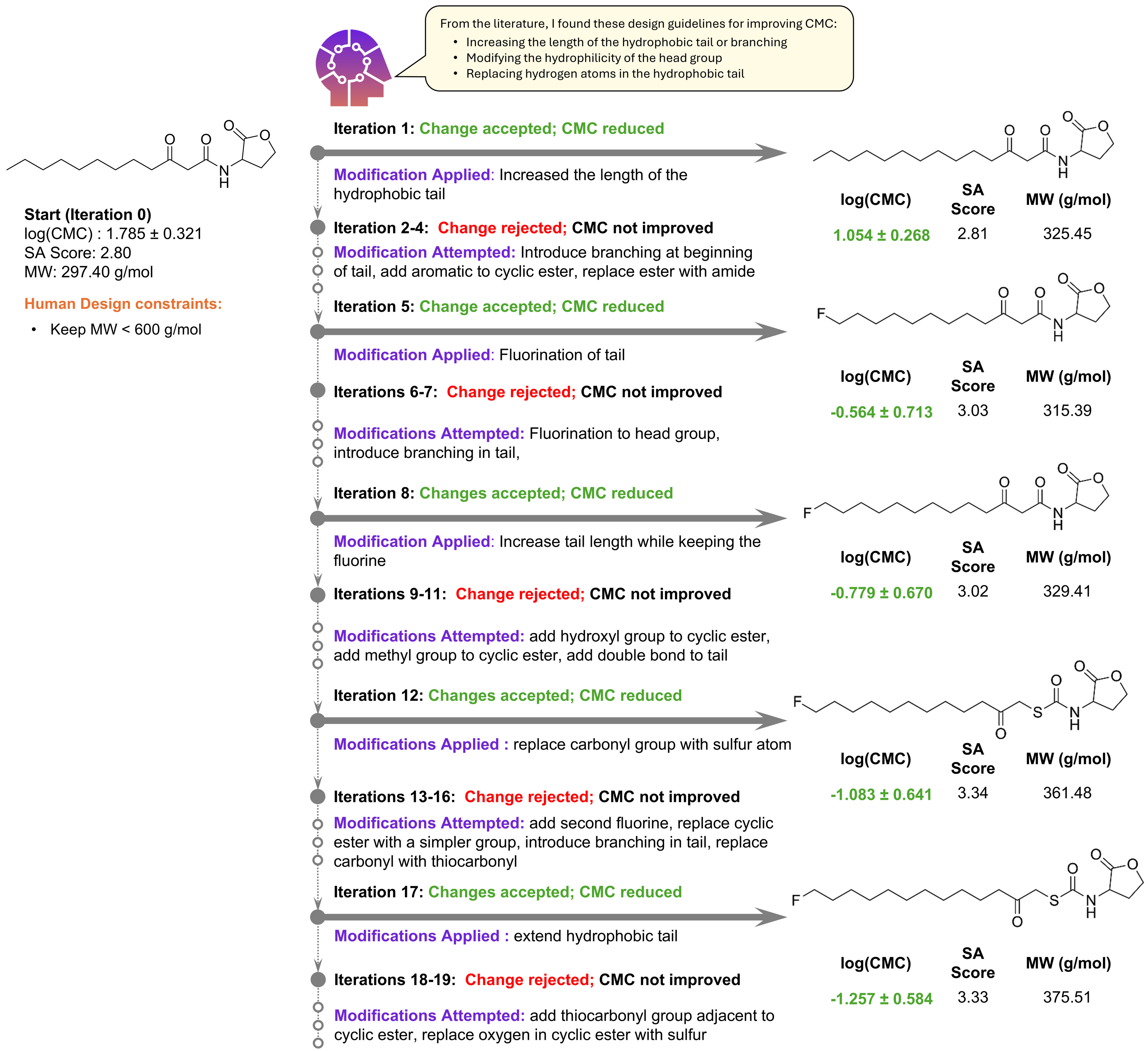}
    \caption{\textbf{dZiner's chain-of-thoughts in the closed-loop inverse design of surfactants with lower CMC}. 
    The agent is powered by Claude 3.5 Sonnet.
    The design guidelines are retrieved from references~\cite{czajka2015surfactants,gaudin2016new,mozrzymas2011prediction,huibers1997prediction,li2004estimation,xuefeng2006correlation,moriarty2023analyzing,boukelkal2024qspr}, and the model is asked to keep the molecular weight lower than 600 (g/mol) in natural language text.
    CMC is reduced by two orders of magnitude via iterative agent-suggested chemical modifications.
 \label{fig:results_CMC}}
\end{figure}
The resulting iterations of surfactant design powered by Claude 3.5 Sonnet (Figure~\ref{fig:results_CMC}) demonstrated the introduction of several modifications to the initial SMILES structure, that ultimately reduced the expected CMC by roughly two orders of magnitude. 
In the resulting benchmarking, Sonnet 3.5 agent generally performed better than the one using GPT-4o, generating a larger proportion of chemically valid surfactant molecules (Table~\ref{tab:benchmark}). 
The associated chain-of-thoughts analysis for GPT-4o generated surfactants is discussed in detail in the supplementary information (see Figure~S1).
In analyzing the Sonnet 3.5 agent generated iterations, the agent was able to initially reduce the log(CMC) by first extending the hydrophobic tail of the initial molecule. After this initial improvement, the agent attempts to make a series of modifications to the head group (iterations 2-5) which are rejected. Iterations 5 and 6 yield the largest reduction to the log(CMC) (improvement by 1.618) with the introduction of fluorine heteroatoms to the tail and head of the surfactant molecule, respectively. Notably, during this heteroatom addition Claude 3.5 Sonnet is able to identify potentially invalid molecular structures due to ambiguity in the learned design rules (add heteroatoms to the end of the tail group), and successfully generates molecules that are valid while applying an equivalent modification (in this case, modifying the terminal carbon on the tail group). 
This behavior was exclusively observed in Claude 3.5 Sonnet  agents, and was not reproduced by GPT-4o, likely contributing to the increased performance of 3.5 Sonnet models.
Another noteworthy behavior is that the suggested changes from iteration 6 were ultimately reverted, as the addition of head group fluorination dramatically increased the SA score, which the model believed was indicative of a synthetic pathway that would not yield further improvement.
Afterwards, additional modifications including the introduction of sulfur heteroatoms and addition groups to the hydrophobic section are able to make modest improvements. Iteration 17 further increased the length of the hydrophobic tail which reduced the log(CMC) to the ultimate value of -1.257, yielding a total improvement of roughly two orders of magnitude over the campaign. The Tanimoto similarity between the initial surfactant molecule and the final molecule was 0.41, demonstrating that for this design task, dZiner is able to make significant and creative changes to the initial molecule over the course of an experiment. 
Throughout the iterations, the SA score ranged from 2.80 to 3.93, where the candidate molecule with the lowest CMC achieved an SA score of 3.33, only slightly more synthetically complex than the initial candidate molecule.
Visualization of the 200 AI-generated molecules in our experiments can be found in Figures S2 and S3 in the Supporting Information.

\subsection{Drug Design and Targeted Docking Inference}
\label{sec:drug}
The discovery of small molecule ligands that bind or inhibit protein targets is ubiquitous to drug development.
However, the design and development of drug candidates is challenging and time-consuming given the multi-objective optimization of biological properties including binding affinity (dissociation constant) (\textit{K$_D$}), solubility, toxicity, and more. Recent advancements in computational methods have focused primarily on de novo discovery~\cite{huang2024dual,zhang2024pocketgen,abramson2024accurate}, while tools to complete hit-to-lead optimization are comparatively lacking. For many novel targets, the discovery and design of small molecules often begins with ligand discovery experiments that discover a ``hit'' molecule with modest binding affinity to the target. 
From this hit, multi-objective optimization must be performed to improve the candidate for further biological study, with significant emphasis placed upon potency or binding affinity. 
Critically, medicinal chemists must be able to synthesize the molecule for testing, somewhat limiting the scope of generative techniques to those that respect synthesizability~\cite{koziarski2024rgfn}.

\begin{figure}[t!]
    \centering
    % \captionsetup{justification=centering}
    \includegraphics[width=0.95\textwidth]{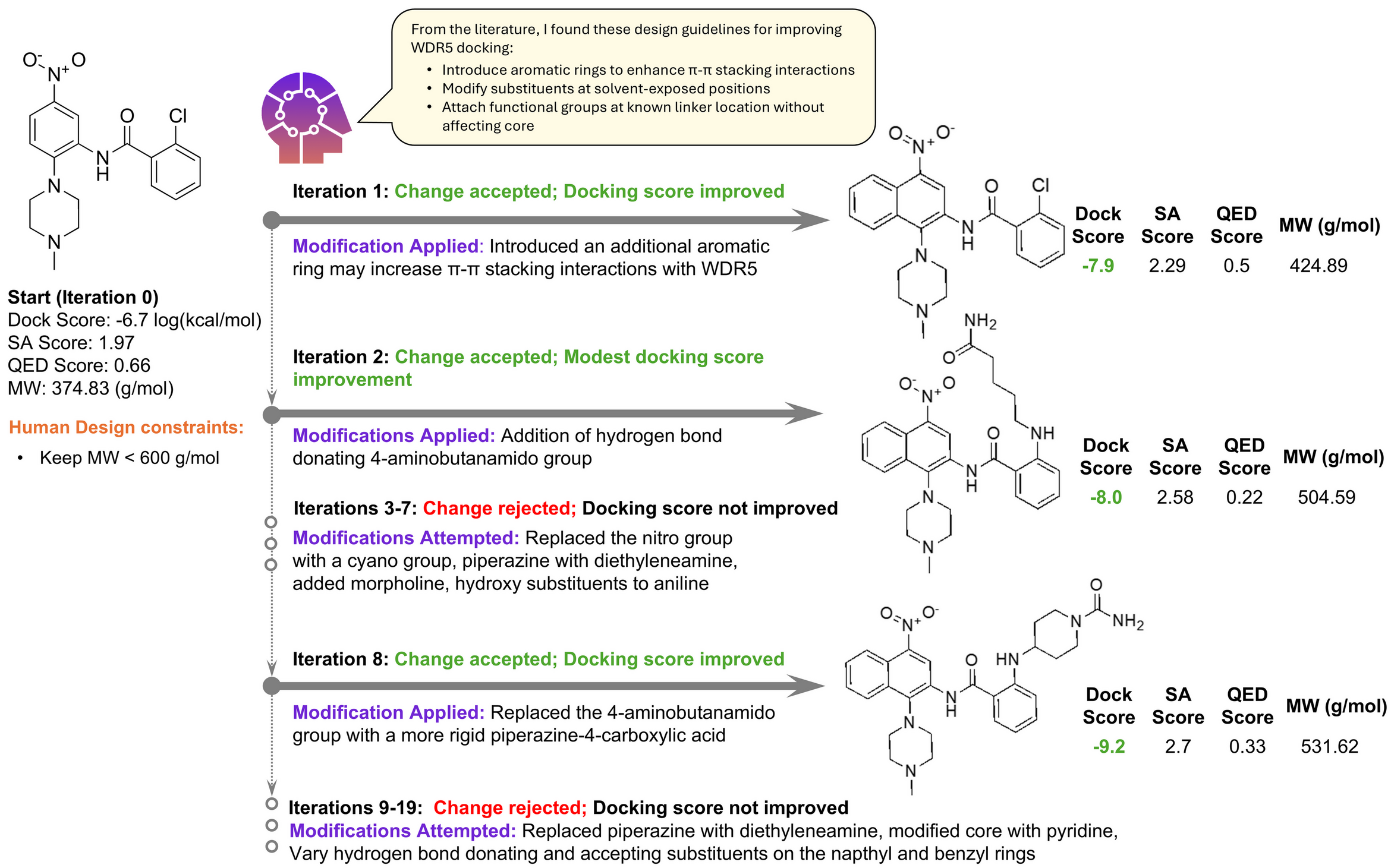}
    \caption{\textbf{
    dZiner's chain-of-thoughts in the closed-loop inverse design of a drug candidate against WDR5 protein target}.
    The design guidelines are extracted by the agent from references~\cite{chen2018design, teuscher2023structure}, and the model is asked to keep the molecular weight lower than 600 (g/mol) in natural language text.
    Docking score is reduced by just over two orders of magnitude via iterative agent-suggested chemical modifications (Dock Score $\propto$ log(kcal/mol)).
 \label{fig:results_docking}}
\end{figure}

Toward this goal, we applied dZiner to the optimization of ligands against WD repeat-containing protein 5 (WDR5). 
WDR5 is a scaffolding protein that plays a critical role in gene expression and cell differentiation through the assembly of chromatin-modifying complexes, such as the MLL/SET methyltransferase complex~\cite{chen2021targeting,grebien2015pharmacological}. 
Thus, WDR5 plays a central role in various cancers by supporting oncogenic transcription. 
%Developing ligands or inhibitors for WDR5 is of interest because disrupting its interactions can inhibit the formation of these transcriptional complexes, offering a potential therapeutic strategy to down regulate oncogene expression and impair cancer cell growth. 
Ligands and inhibitors to WDR5 have been studied and reported, ranging in activity with \textit{K$_D$} and IC$_{50}$ in the 10s of $\mu$M to pM~\cite{aho2019displacement,getlik2016structure}.
This rich background of literature allows the evaluation dZiner's performance, as well as opportunity for human-based input and expertise toward the iterative molecular generation and optimization~\cite{chen2021targeting,getlik2016structure}.
From this literature, we are able to jump into the position of medicinal chemists that have discovered a hit to WDR5 from high-throughput screening (HTS),  demonstrating K$_D$ of 7 ± 1 $\mu$M of the native WIN Peptide substrate (see initial starting molecule in Figure~\ref{fig:results_docking})~\cite{senisterra2013small,getlik2016structure}.
From this HTS hit, dZiner was tasked to perform iterative molecular optimization and improve the binding affinity against WDR5.
Various modifications were applied following key guidelines revealed from literature.
Additionally, the model is asked to keep the molecular weight of the generated candidates lower than 600 (g/mol) in natural language text.
These guidelines, extracted from references~\cite{chen2018design, teuscher2023structure, li2016high}, emphasize the following; 1) enhancing hydrophobic interactions, 2) optimizing solvent-exposed regions, and 3) minimizing steric hindrance without disrupting core binding interactions.
To guide dZiner, we used molecular docking, a common method to approximate the binding affinity of a ligand to a protein.
Though noisy in its recommendation, molecular docking was selected as a surrogate because it is a rapid and computationally inexpensive way to assess the geometric fit of ligands to protein targets and provide an estimation of binding affinity in kcal/mol~\cite{bender2021practical}. 
After each iteration, dZiner docked the generated molecule with WDR5 (PDB: 3UVL) using AutoDock Vina~\cite{trott2010autodock} and a score was computed for each (see Methods Section~\ref{sec:method_docking}). For this task, a reduction in docking score indicates lower binding energy, and thus a higher affinity for WDR5. 

The initial hit provided to dZiner (iteration 0) had a valid structure but relatively modest binding affinity (-6.7). In iteration 1, dZiner adds an aromatic ring to the core to form a rigid napathalene, resulting in a significant improvement in docking score (-7.9). 
In the next iteration, the chloro- substituent on the benzamide is replaced with a bulky, hydrogen bond donating 4-aminobutanamido group, which has a small positive effect on predicted binding affinity. 
In the next iteration, the piperazine ring is replaced with an ethylenediamine functionality, significantly compromising the binding affinity (docking score -6.5). This change is reverted for successive analogs.
In the following iterations, the -NO$_2$ group on the napthyl ring is replaced with an alternative electron-withdrawing group (-CN), a morpholino substituent is added to the benzyl ring, and the benzyl ring itself is swapped for a pyridine.
Each change is found to be detrimental to the docking score and is thus reverted for the next iteration. 

The second substantial improvement in docking score occurs in iteration 8 when the 4-aminobutanamido group is replaced with a more rigid piperidine-4-carboxylic acid. The resulting docking score of -9.2 is significantly better than iteration 0 at -6.7. 
After implementing and reverting several unsuccessful alternatives to the piperazine ring (diethylamine, morpholine), dZiner creates a number of analogs with different H-bond donor and H-bond acceptor groups at variable positions on the benzamide core. 
These minor modifications result in a total of 10 analogs with a docking score $<$ -8.5. 

All iterations produced adhered to the human-provided guideline of MW below 600 g/mol. 
No unstable functional groups were identified in any of the molecules generated by dZiner when using Claude 3.5 Sonnet. 
Given that docking is a low fidelity method of evaluating binding affinity, each of the 10 analogs with docking score $<$ -8.5 would be promising candidates for synthesis in a hit-to-lead campaign.
After 20 iterations, the analogs generated by dZiner have 0.60 Tanimoto similarity to the starting molecule, demonstrating that dziner is capable of making non-trivial modifications to structure to improve binding affinity. 
Furthermore, these designs generally follow those that were used in the development of OICR-9429 (K$_D$ 24 nM)~\cite{grebien2015pharmacological}, maintaining the piperizine ring while adding additional hydrophobic or complementary chemical functionalities~\cite{getlik2016structure}.
When benchmarked against GPT-4o, Claude 3.5 Sonnet generally performed better than GPT-4o, producing fewer invalid SMILES and unstable molecules (see comparison in Table~\ref{tab:benchmark}).
Detailed analysis of the GPT-4o agent generated WDR5 analogs, along with the 200 AI-generated molecules in our experiments can be found in the supplementary information (Figures~S4-6).

\subsubsection{Human-in-the-loop Design}

\begin{figure}[b!]
    \centering
    % \captionsetup{justification=centering}
    \includegraphics[width=\textwidth]{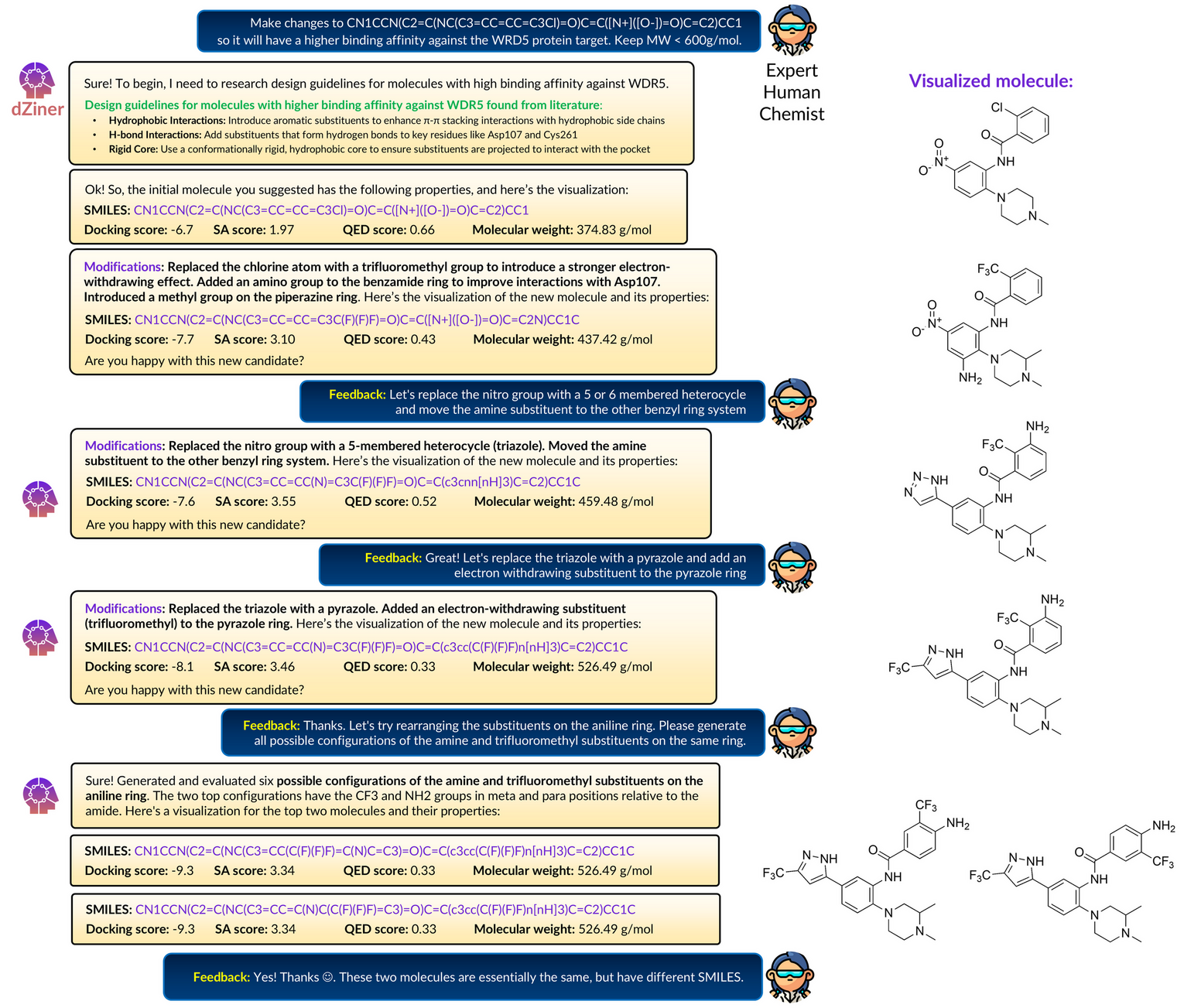}
    \caption{\textbf{Human-in-the-loop inverse design of a drug candidate against WDR5 protein target}.
    The agent is powered by Claude 3.5 Sonnet.
    dZiner found to be highly cooperative, interpretable, and able to enact changes requested with ease in this human-AI collaboration.
 \label{fig:human-in-the-loop}}
\end{figure}

Collaborative efforts between a human expert and AI agents hold significant promise. In the case of molecular design for WDR5 ligands, we examined human guidance to refine the modifications based on docking scores and structural generation (Figure~\ref{fig:human-in-the-loop}). 
Similar to the closed-loop analysis (section~\ref{sec:drug}), the model initially proposed several modifications to the presented structure in accordance with the literature-derived guidelines,
% (with the addition of reference~\cite{li2016high})
 including the addition of an amine group to promote hydrogen bonding with key WDR5 residue Asp107. 
A human chemist reviewed this structure and identified the -NO$_2$ group as a potential toxicophore and priority for replacement. 
Additionally, the human chemist viewed a published crystal structure of WDR5 (41A9) and hypothesized that the -NH$_2$ group should be placed on the other benzyl ring to facilitate interaction with Asp107. 
dZiner was able to competently execute these suggestions and others made by the human chemist to increase binding affinity. dZiner was also able to generate multiple positional isomers when tasked with changing the substitution pattern of an aryl ring. 
The final molecule generated by dZiner and chemist working in tandem had a significantly improved docking score (-9.3) relative to the starting molecule.
Overall dZiner was able to accommodate both general (“add a 5 or 6 carbon heterocycle”) and specific (“do not modify the piperazine”) feedback, and made several creative suggestions for novel WDR5 analogs. 
dZiner was found to be highly cooperative, interpretable, and able to enact most changes requested with ease, even following instructions to revert several iterations and make larger-than-average changes to the molecule.
dZiner was able to effectively generate molecular ligands to WDR5 while managing input-directed goals. Compared to the closed-loop optimization, the human-in-the-loop optimization enabled more diverse and targeted generation (e.g., location of modification) that could be highly beneficial in a variety of contexts to medicinal chemists investigating structure-activity relationships (SAR).
Details on the human-in-the-loop experiment with GPT-4o agent can be found in the supporting information (Figure~S7).
% \rev{Carla please add one or two sentences to compare human-in-the-loop with sonnet 3.5 vs GPT-4o}

\subsection{MOF Organic Linker Design and CO$_2$ Adsorption Capacity Inference\label{sec:results_mof}}
Metal-organic frameworks (MOFs) have gained significant attention in recent years due to their wide range of industrial applications, such as gas adsorption and storage~\cite{li2018recent}, catalysis~\cite{hao2021recent}, and drug delivery~\cite{lawson2021metal}. 
These nanocrystalline porous materials are modular~\cite{kalmutzki2018secondary}, consisting of three main building blocks: inorganic nodes, organic nodes, and organic linkers, along with a specific topology that defines the relative positions and orientations of these building blocks that result different MOF properties. 
\begin{figure}[bh!]
    \centering
    % \captionsetup{justification=centering}
    \includegraphics[width=0.95\textwidth]{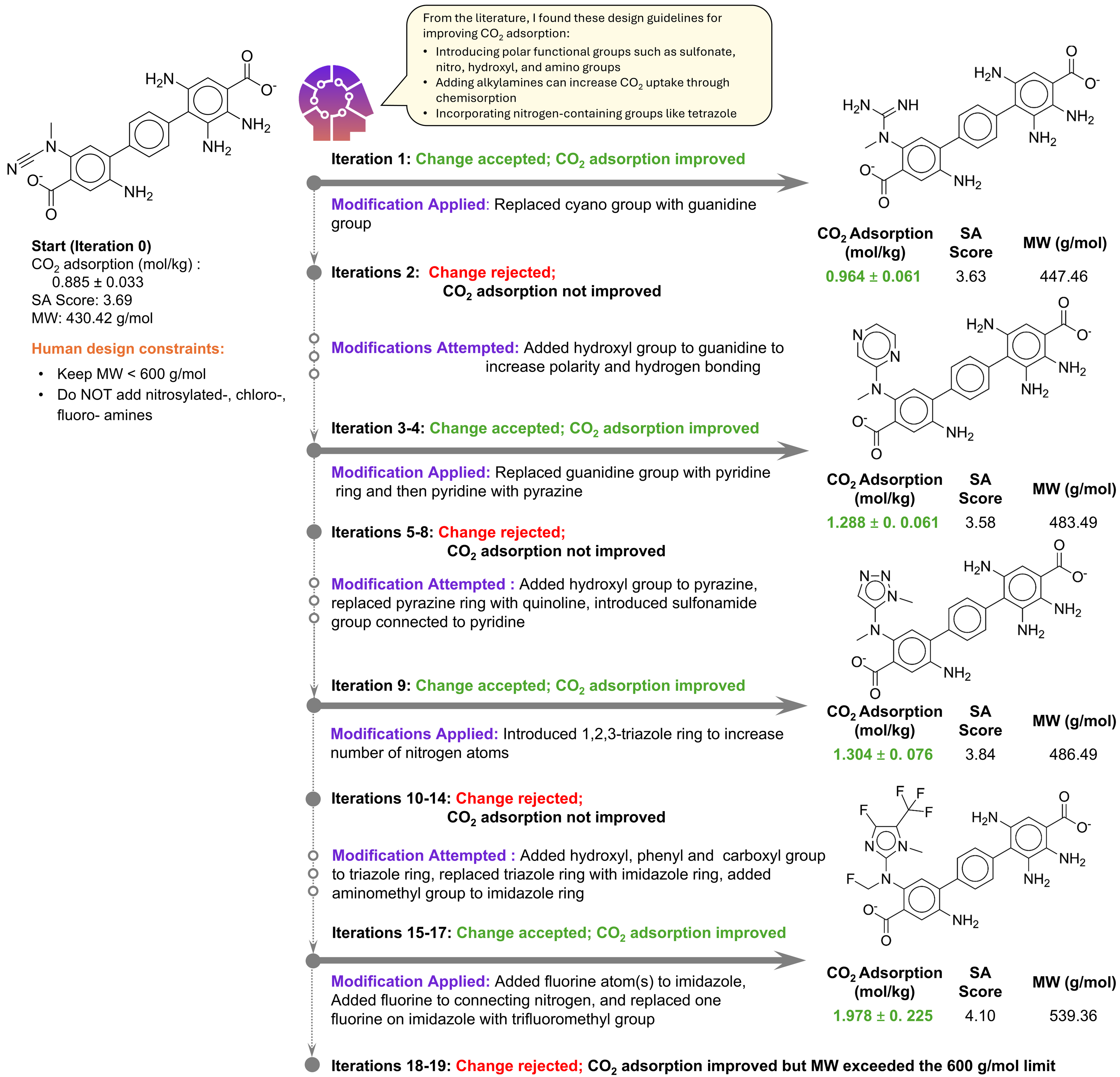}
    \caption{\textbf{
    dZiner's chain-of-thoughts in the closed-loop inverse design of organic linkers for MOFs with high CO$_2$ adsorption capacity}.
    The agent is powered by Claude 3.5 Sonnet.
    Design guidelines were retrieved from references~\cite{usman2022advanced, trickett2017chemistry, matthew2017rational, parveen2024designing}.
    CO$_2$ adsorption capacity is improved by 85\% via iterative agent-suggested chemical
modifications, while following additional design constraints.
 \label{fig:results_CO2}}
\end{figure}
We demonstrate the utility of our framework by applying it to the rational design of likely synthesizable organic linkers for MOFs with high CO$_2$ adsorption capacity at 0.5 bar of pressure. 
These MOFs come with pcu topology and three types of inorganic nodes: Cu paddlewheel, Zn paddlewheel,
and Zn tetramer (three most frequent node-topology pairs in the hMOF dataset~\cite{wilmer2012large}).
In this case, the surrogate CO$_2$ adsorption predictor is an ensemble of fine-tuned MOFormers~\cite{cao2023moformer} trained on extracted SMILES of the organic linkers from their MOFids in the hmof dataset (see section~\ref{methods:mof}).

Throughout the design iterations, as shown in Figure~\ref{fig:results_CO2}, various functional groups were introduced to the initial organic-linker's structure to enhance CO$_2$ adsorption capacity. 
These modifications are derived from guidelines emphasizing the alteration of functional groups, incorporation of nitrogen sites, optimization of pore structures and alkylamines to improve interactions and uptake, as automatically extracted from references~\cite{usman2022advanced, trickett2017chemistry, matthew2017rational, parveen2024designing}.
Additionally, the model includes a set of human design constraints in natural language, specifying to keep the molecular weight lower than 600 (g/mol), and \emph{not} to add nitrosylated, chloro-, fluoro- amines, which are chemically-unstable functional groups (see Figures~S9 and~S10 for the case study without the latter constraint)
The initial organic linker (iteration 0) showed a moderate CO$_2$ adsorption of 0.885. 
In iteration 1, nitrogen-containing functional groups like guanidine were introduced, replacing the cyano group to increase polarity and nitrogen content, which significantly improved CO$_2$ adsorption to 0.964. 
By iteration 3, further modifications involved replacing guanidine with nitrogen-rich heterocycles like pyridine and pyrazine, boosting adsorption to 1.288. 
However, adding polar groups like hydroxyl or carboxyl (iterations 5 and 12) slightly decreased CO$_2$ capture, suggesting that polarity alone was not sufficient for improvement.

The most successful modifications came in iteration 15, where electron-withdrawing fluorine atoms were introduced, leading to a substantial improvement (CO$_2$ adsorption of 1.698), and promoting the electrostatic interactions and hydrogen bonding of the linker.
Further fluorination, including the use of difluoro-substituted and trifluoromethyl groups (iterations 16 to 19), continued to enhance CO$_2$ adsorption, reaching a maximum of 2.289 in iteration 19. 
This increase is attributed to the electron-withdrawing properties of fluorine, which enhance the molecule's interaction with CO$_2$. However, adding too much fluorine also introduced higher uncertainty in adsorption values, indicating potential sensitivity to environmental conditions.
Ultimately, iteration 19 demonstrated that maximizing fluorine content, particularly with trifluoromethyl groups, was the most effective strategy for improving CO$_2$ adsorption.
It is important to note that the last three iterations were not accepted, as the suggested molecules exceeded the 600 g/mol molecular weight limit.
The final accepted linker (iteration 17) showed a significant improvement in CO$_2$ adsorption (1.978) with just 0.50 Tanimoto similarity to the initial molecule, demonstrating that substantial and creative changes were made during the experiment.
Detailed analysis of the GPT-4o agent generated organic linkers, along with the 200 AI-generated molecules in our experiments can be found in the supplementary information (Figures S8, S11 and S12).
% In iteration 1, the introduction of hydroxyl (-OH) and amino (-NH$_2$) groups improved the adsorption to 0.992. 
% However, in iteration 2, the addition of a sulfonate group resulted in a slight decrease in performance.
% Significant improvements were achieved in iteration 7 by incorporating a pyridine ring, which increased nitrogen interactions and boosted CO$_2$ adsorption to 1.278. 
% The highest adsorption, 1.644, was observed in iteration 8 when a fluorine atom was introduced, leveraging its high electronegativity to enhance CO$_2$ capture. 
% A chlorine atom was also added in iteration 9, resulting in a CO$_2$ adsorption of 1.409. 
% Overall, the combination of electronegative atoms and nitrogen-containing functional groups proved most effective in enhancing CO$_2$ adsorption.
% Throughout the optimization, the molecular weight increased from 430.424 g/mol (iteration 0) to 516.945 g/mol (iteration 9). The SA score also fluctuated, peaking at 4.595 in iteration 5 after the addition of hydroxyl groups, indicating increased synthetic complexity.
% \subsection{Peptide Design and Hemolytic Activity Inference}
\section{Discussion}
\label{sec:discission}
Our workflow, dZiner, represents an agent-based computational framework for accelerated materials discovery by replicating and incorporating the expertise of human domain experts across various inverse design tasks and target properties, including surfactants, ligand and drug design, and metal-organic frameworks.
The inclusion of other expert tools is easy, and our examples demonstrate that dZiner is generally able to adapt across various property-to-structure problems in materials discovery.
To better assess the impact of domain-knowledge in our experiments, we repeated all three case studies by removing the design guidelines retrieval tool from the scientific literature.
This served as the baseline.
In this setting, the agent's modifications to the core structure of the molecules were primarily restricted to the addition or removal of random functional groups and elements, due to the limited non-domain-specific knowledge of the stand-alone LLM at the training time.

Tables~\ref{tab:benchmark} and~\ref{tab:benchmark-improvement} provide a detailed breakdown of dZiner's performance across three inverse design tasks—CMC, WDR5 docking, and CO$_2$ adsorption—evaluating its success with and without domain-knowledge (each was used to generate 600 molecules across all tasks).
\begin{table}[!th]
\centering
\caption{dZiner's success rates for the three inverse design tasks with different target properties, evaluated over the generation of 100 molecules per task.
Primary objectives for each task are in bold. 
The baselines include model runs without retrieving domain-knowledge (design guidelines) from the scientific literature.\label{tab:benchmark}}
\resizebox{\textwidth}{!}{%
\begin{tabular}{@{}lcccccc@{}}
\toprule
\textbf{Target Property}               & \textbf{Criteria}                                             & \multicolumn{2}{c}{\textbf{Success Rate without Domain-knowledge (\%)}} & \multicolumn{2}{c}{\textbf{Success Rate with Domain-knowledge (\%)}} \\ 
                            &                                                              & \textbf{GPT-4o (baseline)} & \textbf{Claude 3.5 Sonnet (baseline)} & \textbf{GPT-4o} & \textbf{Claude 3.5 Sonnet} \\ \midrule
\textbf{Task 1: CMC}        & Valid SMILES Generation                                       & 77   & 89   & 79   & 96 \\
                        
                            & \textbf{Lower log(CMC)}                                         & 55   & 81   & 91   & 92 \\
                            % & Lower MW                                   &     &     & 26   & 8  \\
                            & Meeting MW Design Constraint               & 100   & 100   & 95   & 100 \\
                            & Lower SA Score                     & 31  & 24   & 23   & 19 \\ \midrule
\textbf{Task 2: WDR5 Docking} & Valid SMILES Generation                                      & 63  & 96   & 83   & 100 \\
                        
                            & \textbf{Lower Docking Score}                              & 60   & 89   & 81   & 96 \\
                            % & Lower MW                               &     &     & 20   & 7  \\
                            & Meeting MW Design Constraint               & 100   & 100   & 99   & 100 \\
                            & Lower SA Score                     & 43   & 19    & 22   & 6  \\
                            & Higher QED Score     & 47   & 46   & 13   & 18 \\ \midrule
\textbf{Task 3: CO$_2$ Adsorption} & Valid SMILES Generation                                   & 44   & 99   & 86   & 100 \\
                          
                            & \textbf{Higher CO$_2$ Adsorption}                             & 39   & 98   & 77   & 98 \\
                            % & Lower MW                                   &     &     & 24   & 5  \\
                            & Meeting MW Design Constraint               & 97   & 97   & 97   & 95 \\
                            & Lower SA Score                     & 76   & 98   & 63   & 41   \\ \bottomrule
\end{tabular}%
}
\end{table}
Across these tasks, dZiner, powered by Claude 3.5 Sonnet, outperforms GPT-4o significantly in both conditions, with especially high success rates when leveraging domain-knowledge from the literature. 
This is notable in primary objectives such as improving log(CMC), binding affinity (docking score), and CO$_2$ adsorption.
On the importance of incorporating literature-gathered and human expert-based design principles in the workflow, the baseline runs, which operated without design guidelines, exhibited a high failure rate in terms of both generating valid molecular structures and optimizing the target properties, regardless of the choice of the LLM. 

We quantified the success of the model in meeting primary objectives by 1) assessing the average improvement in log(CMC), docking score, CO$_2$ adsorption (Table~\ref{tab:benchmark-improvement}) for the best candidate in each run; 2) by comparing each generated iteration to the initial candidate (iteration 0) to see what percentage of molecules have improved on the target property (Table~\ref{tab:benchmark}). 
GPT-4o struggled to generate valid SMILES and to suggest new molecules with improved primary objectives. 
Claude 3.5 Sonnet consistently performed better than GPT-4o on this metric.
% GPT-4o, in particular, struggled to generate valid SMILES and suggest new molecules with substantially better (average improvement)-or even slightly better (see lower log(CMC), lower docking score and higher CO$_2$ adsorption criteria in Table~\ref{tab:benchmark}) target properties than the initial candidates at iteration 0. 
For instance, its success rate in valid SMILES generation for CMC and CO$_2$ adsorption tasks was considerably lower than Claude 3.5 Sonnet, highlighting its limitations in molecule design tasks without specialized guidance.
On the other hand, it adhered to the human design constraints (molecular weight and the choice of forbidden functional groups), while still optimizing target properties.
This efficiency underscores the value of literature-gathered and human expert-based design principles, in complex molecular design tasks, where balancing various criteria is essential for overall success.

\begin{table}[!b]
\centering
\caption{dZiner's improvement rates for the primary objectives in the three inverse design tasks with different target properties, evaluated over the generation of 100 molecules per task.
On average, improvements are determined by comparing the best candidate from each run to the initial candidate (iteration 0).
The baselines include model runs without retrieving domain-knowledge (design guidelines) from the scientific literature.\label{tab:benchmark-improvement}}
\resizebox{\textwidth}{!}{%
\begin{tabular}{@{}lcccccc@{}}
\toprule
\textbf{Target Property}               & \textbf{Criteria}                                             & \multicolumn{2}{c}{\textbf{Improvement without Domain-knowledge (\%)}} & \multicolumn{2}{c}{\textbf{Improvement with Domain-knowledge (\%)}} \\ 
                            &                                                              & \textbf{GPT-4o (baseline)} & \textbf{Claude 3.5 Sonnet (baseline)} & \textbf{GPT-4o} & \textbf{Claude 3.5 Sonnet} \\ \midrule
\textbf{Task 1: CMC}        &\\
                            & \textbf{Average log(CMC)}                                        & 34   & 86   & 95   & 137 \\
                            \\ \midrule
\textbf{Task 2: WDR5 Docking} & \\
                            & \textbf{Average Docking Score}                                        & 16  & 19   & 31   & 31 \\
                           \\ \midrule
\textbf{Task 3: CO$_2$ Adsorption} & \\
                             & \textbf{Average CO$_2$ Adsorption}                                        & 41   & 28   & 46  & 108 \\
 \\ \bottomrule
\end{tabular}%
}
\end{table}

 % Across these tasks, dZiner, powered by Claude 3.5 Sonnet, outperforms GPT-4o significantly in both conditions, with especially high success rates when leveraging domain-knowledge from the literature. This is notable in primary objectives such as improving LogCMC, binding affinity (docking score), and CO$_2$ adsorption.

% For the primary objectives (lower CMC, higher binding affinity, and higher CO$_2$ adsorption), dZiner was generally successful.
% dZiner efficiently modified the molecules across all tasks, generating very few invalid structures while adhering to design guidelines from the literature and human-provided criteria. 
% This included maintaining a molecular weight below 600 g/mol, a crucial factor for ensuring molecular efficiency. 
% In particular, larger molecular weight compounds typically experience improved molecular docking scores, meaning that efficiency must be accounted for in the optimization.

Our approach offers several key contributions;
% \begin{itemize}
1) The model's flexibility enables the integration of the complete property optimization task with additional design constraints directly through natural language, making the workflow easily adaptable to different target properties simply by altering the input query (prompt).
2) The augmented domain-expert surrogate models can be easily customized to target specific properties. 
This flexibility allows users to either train their own machine learning or deep learning models, or better yet, leverage the existing \emph{state-of-the-art} property predictors from the materials community, avoiding the unnecessary effort of reinventing the wheel.
This approach also opens up the possibility of uncertainty estimations of the predicted property via an ensemble of inference models, a capability that typical standalone LLMs do not possess.
3) The model provides \emph{chain-of-thought reasoning}, enabling more interpretable results and a clearer understanding of its chemistry-informed decision-making processes.
4) The workflow supports both closed-loop and human-in-the-loop inverse design.
In the human-in-the-loop scenario, a domain expert can interact with the model through natural language to provide feedback on newly suggested candidates, propose modifications, or introduce additional design constraints.
5) Because of the iterative design approach, we observed that most molecular candidates maintained a strong relative amount of synthesizbility compared to completely generative approaches, especially seen in the CMC and WDR5 ligand design.

\subsection{Limitations and Future Work}
As demonstrated in the previous sections, dZiner works well as an engine for accelerated molecular discovery and in silico experimentation across a wide range of chemical tasks.
 Despite this flexibility, there are still areas where dZiner could be further improved.
 The incorporation of multi-modal data (such as the ability to interpret images and schemes) from supporting literature represents a key advancement that could further improve the performance of the model. 
This improvement would be especially impactful for small molecule virtual screening tasks. 
% In addition, the current framework leverages SMILES and is thus limited to molecular design tasks that are well represented by SMILES.
% For stochastic design tasks (such as the synthesis of polymers) the model would require extension to other molecular representations (e.g. BIGSMILES, SMARTS, etc).
% The modeinherests the limitations of SMILES as a molecular representation.
For example, SMILES can be a major
oversimplification, especially for complex structures like MOFs.
Another limitation inherent to SMILES is that unique SMILES codes can be generated for each substitution pattern, leading to a lack of a one-to-one mapping between molecules and their SMILES representations.
These limitations may be balanced by considering that SMILES are likely the most prevalent molecular representations available in the training data of LLMs, making it easier for LLMs to successfully generate valid new candidates.
While the current framework leverages textual representations of molecules, it is not necessarily limited to this format, and the agent can easily incorporate other representations.
For example, for stochastic design tasks such as the synthesis of polymers, molecular representations like BigSMILES~\cite{lin2019bigsmiles} or SMILES Arbitrary Target Specification (SMARTS)~\cite{daylight2007smarts}, which may better capture the complexity of such systems~\cite{stuart2023sizing}. It should be noted that given that these other molecular representations are not as popular, the model may lose performance when generating valid candidates.
For human-in-the-loop runs, the model had limited success in interpreting terms like ortho, meta, and para when a ring has three or more substituents. 
As an example, it struggled with prompts like ``move the -NH$_2$ group so it is para to the -CF$_3$ group''. 
This likely a result of the zero-shot learning approach and could be resolved with further domain-specific fine-tuning. Alternatively, adding more instructive human feedbacks or augmenting the agent with additional validation tools to better incorporate chemical rules is expected to alleviate this limitation.
% \begin{itemize}
%       \item The model's flexibility enables the integration of the complete property optimization task with additional design constraints directly through natural language, making the workflow easily adaptable to different target properties simply by altering the input query (prompt). 
%       \item The augmented domain-expert surrogate models can be easily customized to target specific properties. 
%       This flexibility allows users to either train their own machine learning or deep learning models, or better yet, leverage the existing \emph{state-of-the-art} property predictors from the materials community, avoiding the unnecessary effort of reinventing the wheel.
%       This approach also opens up the possibility of uncertainty estimations of the predicted property via an ensemble of inference models, a capability that typical standalone LLMs do not possess.
%       \item The model provides \emph{chain-of-thought reasoning}, enabling more interpretable results and a clearer understanding of its chemistry-informed decision-making processes.
%       \item The workflow supports both closed-loop and human-in-the-loop inverse design.
%       In the human-in-the-loop scenario, a domain expert can interact with the model through natural language to provide feedback on newly suggested candidates, propose modifications, or introduce additional design constraints.
% \end{itemize}
\section{Methods}
\label{sec:methods}
\subsection{AI Agent}
In broader terms, an \emph{agent} refers to an entity capable of taking action.
The AI agent in this work is powered by an LLM, acting as its brain, expanding its perceptual and action space (environment) through strategies such as multimodal perception and tool utilization~\cite{nakano2021webgpt,yao2022react, schick2023toolformer, lu2023chameleon, qin2023tool}. 
In this work, we exploit the \emph{zero-shot learning} capabilities of large language models (LLMs)\cite{kojima2022large} alongside the ReAct architecture, which supports both reasoning (e.g., chain-of-thought prompting) and taking actions (e.g., generating action plans)\cite{yao2022react}.
Reasoning traces guide the model in creating, overseeing, and refining action plans, while also addressing exceptions. 
At the same time, actions enable the model to interact with external sources like knowledge bases or environments to acquire additional information. 
These knowledge bases are structured as toolkits (see section~\ref{sec:toolkits}), enabling the agent to extract relevant molecular design insights from research papers, publicly available datasets, and advanced built-in chemical knowledge, as well as on-the-fly evaluation of modifications with domain-expert surrogate models and synthesizability assessment (Algorithm~\ref{dziner_alg}).
This specialized materials design guidelines lie beyond the scope of typical LLM's training data, enhancing the model’s ability to function as an expert chemist in various domains in a zero-shot manner. In this work, we used OpenAI’s GPT-4o~\cite{openai2024hello} and Anthropic's Claude 3.5 Sonnet~\cite{claude_sonnet_35} with a temperature of 0.3 as our agent's LLM, and LangChain~\cite{chase2022langchain} for the application framework development.

\begin{algorithm}[t!]
\caption{dZiner Algorithm (closed-loop)\label{dziner_alg}}
\begin{algorithmic}
    \State \textbf{Input:} $x_0$: chemical structure of the starting molecule
    \State \hspace{2.7em} $y_{\text{target}}$: target property label to optimize (can be a minimization or maximization problem).
    \State \hspace{2.7em} $\mathcal{L}$: selection of scientific literature on target property optimization.
    \State \hspace{2.7em} $\mathcal{S}(x)$: surrogate domain-expert model for evaluating target property, given $x$.
    \State \hspace{2.7em} $\mathcal{M} := \emptyset$: set of history messages, if any.
    \State \hspace{2.7em} $(x_{\text{best}}, \hat{y}_{\text{best}}, \hat{\sigma}_{\text{epi}_{best}}^2) \gets (x_0, (\hat{y}_0, \hat{\sigma}_{\text{epi}_{0}}^2):\mathcal{S}(x_0))$: initialize best molecule.
    \State \textbf{Output:} $(x_i, y_i, \hat{\sigma}_{\text{epi}_{i}}^2)$: chemical structure and property of the new molecule.
    
    \For{$i = 1:N$}
        \State $m_i, g_i \gets \text{LLM}(x_{i-1}, \hat{y}_{i-1}, y_{\text{target}}, \mathcal{L}, \mathcal{M})$ \Comment{$m$: modification, $g$: guideline, $\mathcal{M}$: history}
        \State $\tilde{x}_i \gets \text{LLM: modify structure}(x_{i-1}, g_i)$ \Comment{$\tilde{x}_i$: modified molecule}
        \State $x_i , v_i \gets \text{$\mathcal{V}(\tilde{x}_i$)}$ \Comment{$x_i$: new molecule, $v_i$: validity check}
        
        \If{ $v_i$ is invalid}
            \State \textbf{return} Invalid molecule
            \State $x_i, \hat{y}_i, \hat{\sigma}_{\text{epi}_{i}}^2 \gets x_{\text{best}}, \hat{y}_{\text{best}}, \hat{\sigma}_{\text{epi}_{best}}^2$ \Comment{Revert to best}
            \State \textbf{continue}
        \EndIf
        
        \State $\hat{y}_i, \hat{\sigma}_{\text{epi}_i}^2 \gets \text{$\mathcal{S}(x_i$)}$ \Comment{$\hat{y}_i$: property, $\hat{\sigma}_{\text{epi}_{i}}^2$: epistemic uncertainty}
        
        \If{$|\hat{y}_i - y_{\text{target}}| \geq |\hat{y}_{\text{best}} - y_{\text{target}}|$}
            \State $x_i, \hat{y}_i, \hat{\sigma}_{\text{epi}_i}^2 \gets x_{\text{best}}, \hat{y}_{\text{best}}, \hat{\sigma}_{\text{epi}_{best}}^2$ \Comment{Revert to best}
            \State \textbf{continue}
        \Else
            \State $(x_{\text{best}}, \hat{y}_{\text{best}}, \hat{\sigma}_{\text{epi}_{best}}^2) \gets (x_i, \hat{y}_i, \hat{\sigma}_{\text{epi}_{i}}^2)$ \Comment{Update best}
        \EndIf
        
        \State $h_i \gets \text{create history message}(x_i, \hat{y}_i, \hat{\sigma}_{\text{epi}_{i}}^2, v_i, g_i)$ \Comment{$h_i$: history message}
        \State $\mathcal{M} \gets \mathcal{M} \cup \{h_i\}$ \Comment{Update history}
    \EndFor
    \Statex
    \State \textbf{Notation:}
    \State \hspace{1.5em} $N$: max number of new molecules
    \State \hspace{1.5em} $\mathcal{V}$: Chemical feasibility and synthesizeability assessment; $\mathcal{M}$: history
\end{algorithmic}
\end{algorithm}

\subsubsection{Agent Toolkits}
\label{sec:toolkits}
\subsubsection{Domain-expert Knowledge}
This tool enables the agent to do retrieval-augmented generation (RAG), and extract design guidelines from unstructured text, offering insights on how to modify the core structure of a molecule to optimize a specific property. 
It identifies the most relevant sentences from research papers in response to a query, focusing on suggestions for molecular modifications that enhance the desired property. 
The process involves embedding both the paper and the query as numerical vectors, and then selecting the top $k$ passages within the document that either explicitly mention or implicitly hint at adaptations to optimize the band gap property of a MOF. The embedding model used is OpenAI’s text-embedding-3-large.
Drawing on our previous work~\cite{ansari2023agent}, $k$ is set to 9 but is dynamically adjusted based on the context’s length to prevent OpenAI’s token limitation errors.
The semantic similarity search is ranked using Maximum Marginal Relevance (MMR)~\cite{carbonell1998use}, based on cosine similarity, which is defined as:
\begin{equation}
\text{MMR} = \arg\max_{d_i \in R \setminus S} \left[ \lambda \cdot \cos(d_i, q) - (1 - \lambda) \cdot \max_{d_j \in S} \cos(d_i, d_j) \right]
\label{eq:MMR}
\end{equation}
Here, $d_i$ represents a document from the set of retrieved documents $R$, $S$ is the set of already selected documents, and $q$ is the query. The parameter $\lambda$, which ranges from 0 and 1, controls the balance between relevance to the query and diversity (i.e., novelty compared to the already selected documents). In this work, we use the default value of 0.5. The purpose of MMR is to retrieve documents that are both relevant to the query and diverse, minimizing redundancy in the results.

\subsection{Domain-expert Surrogate Models}
For the case studies in sections~\ref{sec:results_CMC} and~\ref{sec:results_mof}, we use ensemble modeling to estimate prediction uncertainty, thus enhancing the predictive capability of the domain-expert surrogate model.
For a given data point $\vec{x}$, the ensemble prediction average ($\hat{y}(\vec{x})$) is calculated as follows:
\begin{equation}
\hat{y}(\vec{x}) = \frac{1}{N} \sum_m \hat{y}_m(\vec{x})
\label{eq:uncertainty}
\end{equation}

\begin{equation}
\hat{\sigma}_{\text{epi}}^2(\vec{x}) = \frac{1}{N} \sum_m \left( \hat{y}(\vec{x}) - \hat{y}_m(\vec{x}) \right)^2
\label{eq:uncertainty}
\end{equation}
where $N$ is the ensemble size, and $m$ indexes the model in the ensemble. 
$\hat{\sigma}_{\text{epi}}^2(\vec{x})$ denotes the epistemic uncertainty, quantifying the disagreements amongst model estimations.

\subsubsection{ Critical Micelle Concentration Inference Model}
\label{methods:cmc}
This model is based on the work of~\citet{qin2021predicting} without any adaptations.
The authors used a Graph Convolutional Neural Network (GCN) to predict the critical micelle concentration (CMC) of surfactants based on their molecular structure as SMILES input. 
The CMC training data were experimentally measured at room
temperature (between 20 and 25 $^{\circ}$C) in water for 202 surfactants coming from various classes, including nonionic, cationic, anionic, and zwitterionic.
The model architecture leverages graph convolutional layers to process molecular graphs, where atoms are represented as nodes and bonds as edges, effectively capturing both topological and constitutional information. 
The GCN includes average pooling to aggregate atom-level features into a fixed-size graph-level vector, followed by fully connected layers with ReLU activations for the final regression of the log CMC value. 
In terms of performance, the GCN has a root-mean-squared-error (RMSE) of 0.23 and an $R^2$ of 0.96 on test data for nonionic surfactants, outperforming previous quantitative structure-property relationship (QSPR) models. The ensemble of models used in our study are based on an 11-fold cross-validation with mean RMSE of 0.32.
For more details on the model, refer to reference~\cite{qin2021predicting}.

\subsubsection{Targeted Docking Inference Model}
\label{sec:method_docking}
This model utilizes on Dockstring~\cite{garcia2022dockstring}, to predict the fit and binding affinity of small molecules (ligands) bind to target proteins by using molecular docking. 
Dockstring is a user-friendly Python wrapper for AutoDock Vina~\cite{trott2010autodock}. 
WDR5 (PDB: 3UVL)~\cite{zhang2012plasticity} was accessed on May 30th 2024 and was prepared for molecular docking in MGLTools / Python Molecular Viewer (1.5.7) by removing the Histone-lysine N-methyltransferase MLL3 peptide ligand, cofactors, and water. The protein was protonated at pH 7.4 by adding Polar Only Hydrogens. Kollmann charges were added. Dockstring uses a targeted version as opposed to blind docking and a 30 Angstrom grid box was defined around the central binding pocket of WDR5. Docking was performed using default exhaustiveness and energy range. Docking results were returned and used without any rescoring in single measurements. 

\subsubsection{CO$_2$ Adsorption Inference Model}
\label{methods:mof}
This model is based on MOFormer, a self-supervised Transformer model developed for predicting the properties of Metal-Organic Frameworks (MOFs) using a structure-agnostic approach~\cite{cao2023moformer}. 
Unlike traditional models that rely on 3D atomic structures, MOFormer uses a text-based representation of MOFs, known as MOFid. The model utilizes the self-attention mechanism of Transformers to capture complex relationships within MOFs and is pretrained on over 400,000 MOF structures using self-supervised learning with Barlow-Twin loss~\cite{zbontar2021barlow}. 
This pretraining improves the prediction accuracy, as it aligns the textual-based representations of MOFormer with the structure-based representation leaning of a Crystal Graph Convolutional Neural Network (CGCNN)~\cite{xie2018crystal}.
For the specific task of predicting CO2 adsorption capacity at 0.5 bar, MOFormer achieved a mean absolute error (MAE) of 0.545 mol/kg, whereas our fine-tuned ensemble of models with 5-fold training on the SMILES of the organic linkers have a MAE of 0.894  mol/kg.
For more details on the model, refer to reference~\cite{cao2023moformer}.

\subsection{Synthesizability Assessment}
This tool uses RDKit~\cite{landrum2013rdkit} to convert a SMILES string into an RDKit \emph{Mol} object and performs several validation steps, including syntax parsing, atom and bond validation, checking atomic valences, verifying ring closure notation, adding implicit hydrogens, and detecting aromatic systems. These processes ensure the basic chemical validity of the molecule.
In addition to the chemical feasibility assessment, we use a heuristic measure of synthesizability: synthetic accessibility
score (SA score)~\cite{ertl2009estimation}, which is
based on the analysis of one million PubChem molecules and combines fragment contributions from molecular substructures with a complexity penalty that accounts for molecular size and structural features.
In section~\ref{sec:drug}, we also include a quantitative estimate of
drug-likeness (QED)~\cite{bickerton2012quantifying}, which measures a compound's drug-likeness by integrating molecular properties, such as molecular weight, lipophilicity (logP), polar surface area, and the number of hydrogen bond donors and acceptors, into a single value.
% \subsubsection{Peptide Hemolytic Activity Inference Model}
\section*{Data and Code Availability}
All data, code and model architectures and fine-tuned weights for the surrogate models used to produce results in this study are publicly available in the following GitHub repository:
\href{https://github.com/mehradans92/dZiner}{\color{blue}{{https://github.com/mehradans92/dZiner}}}.
\section*{Acknowledgments}
This research was undertaken thanks in part to funding provided to the University of Toronto's Acceleration Consortium from the Canada First Research Excellence Fund: Grant number - CFREF-2022-00042.
The authors thank Santha Santhakumar at the Acceleration Consortium for his valuable feedback on the chemical assessment of the AI-generated molecules. 
\bibliographystyle{unsrtnat}
\bibliography{bibliography}

% \section{Supplementary Information}
% \label{sec:SI}

% making it a promising tool for high-throughput screening, particularly when the 3D structure of MOFs is not readily available.

% \subsection{Supplementary Figures}

\end{document}

\maketitle

% \section*{Supporting figures}

\renewcommand{\thefigure}{S\arabic{figure}}
\setcounter{figure}{0}

\section{Surfactant Design and Critical Micelle Concentration Inference with GPT-4o\label{sec:si_results_CMC}}

\begin{figure}[H]
    \centering
    % \captionsetup{justification=centering}
        \includegraphics[width=0.95\textwidth]{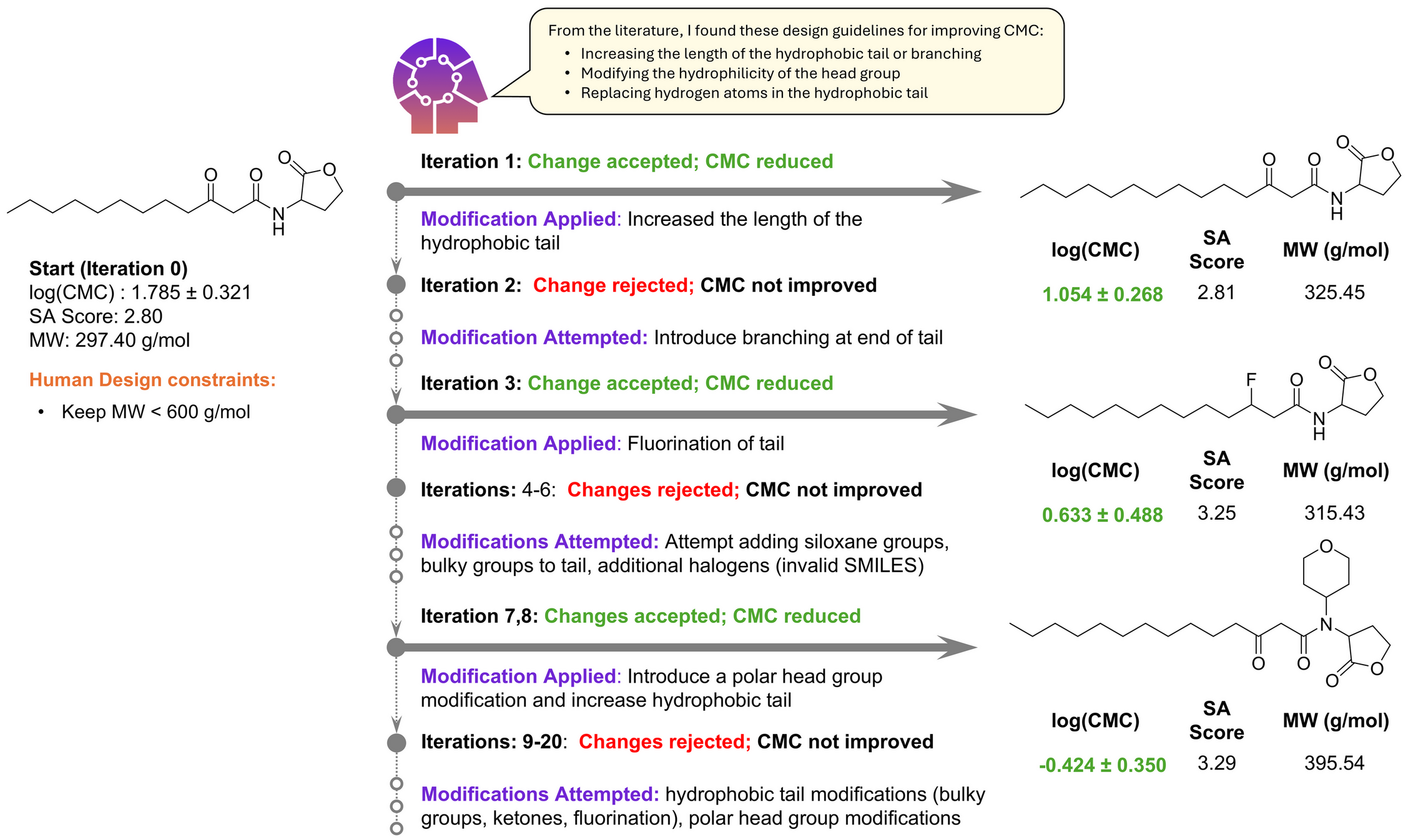}
    \caption{\textbf{dZiner's chain-of-thoughts in the closed-loop inverse design of surfactants with lower CMC}. 
    The agent is powered by GPT-4o.
    The design guidelines are retrieved from literature (same references as in Figure 2), and the model is asked to keep the molecular weight lower than 600 (g/mol) in natural language text.
    CMC is reduced by two orders of magnitude via iterative agent-suggested chemical modifications. The accepted molecule bears 0.74 similarity (Tanimoto) to the starting molecule after 8 iterations.
 \label{fig:si_results_CMC_SI}}
\end{figure}

Similar to Section 2.1, and starting from the same initial surfactant molecule, we applied dZiner powered by GPT-4o to this property optimization task.
The resulting iterations of surfactant design (Figure~\ref{fig:si_results_CMC_SI}) demonstrated the introduction of several modifications to the initial SMILES structure, that ultimately reduced the expected CMC by roughly two orders of magnitude. 
Across the first 3 iterations of design, the agent was able to significantly reduce the CMC by introducing additional methyl-type units to the hydrophobic tail (iteration 1), as well as replacing hydrogen atoms in the tail with fluorine (iteration 3). 

During this improvement, iteration 2 attempted to introduce branching in the hydrophobic tail, but was rejected after the CMC evaluation did not yield any improvement between iterations 1 and 2. 
Following several iterations with other rational but ultimately unsuccessful modifications, the agent achieves the largest reduction of log(CMC) in iteration 7 (0.633 to 0.102) by replacing the head group with a series of amide-linked cyclic ethers.
Interestingly, the agent completes this modification at the expense of the modification in iteration 3, which ultimately further reduces the CMC beyond what was previously achieved (this behavior also occurred in other benchmarking runs). 
The final improvement in log(CMC) was achieved in iteration 8 with a further increase to the length of the hydrophobic tail unit to the ultimate value of -0.424. 
Throughout the experiments the SA score ranged from 2.80 to 4.01, where the candidate molecule with the lowest CMC achieved an SA score of 3.29, only slightly more complex than the initial candidate molecule.

\begin{figure}[H]
    \centering
    % \captionsetup{justification=centering}
    \includegraphics[width=\textwidth]{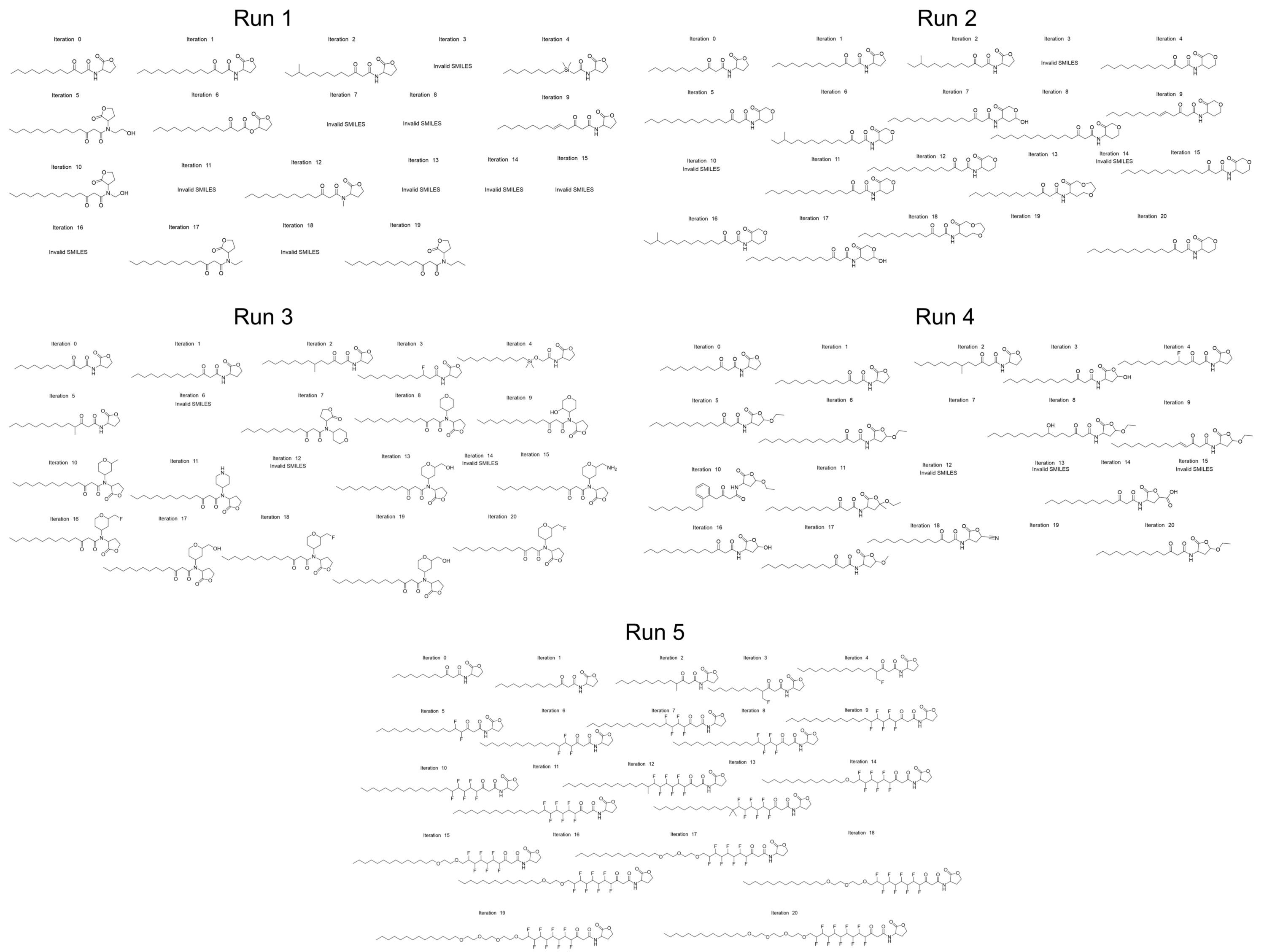}
    \caption{Visualization of the 100 molecules generated in the closed-loop inverse design of surfactants with lower CMC. The agent is powered by GPT-4o.
    No potentially unstable functional groups were found.
    Invalid SMILES generated are marked as invalid.
 \label{fig:cmc_SI}}
\end{figure}

\begin{figure}[H]
    \centering
    % \captionsetup{justification=centering}
    \includegraphics[width=\textwidth]{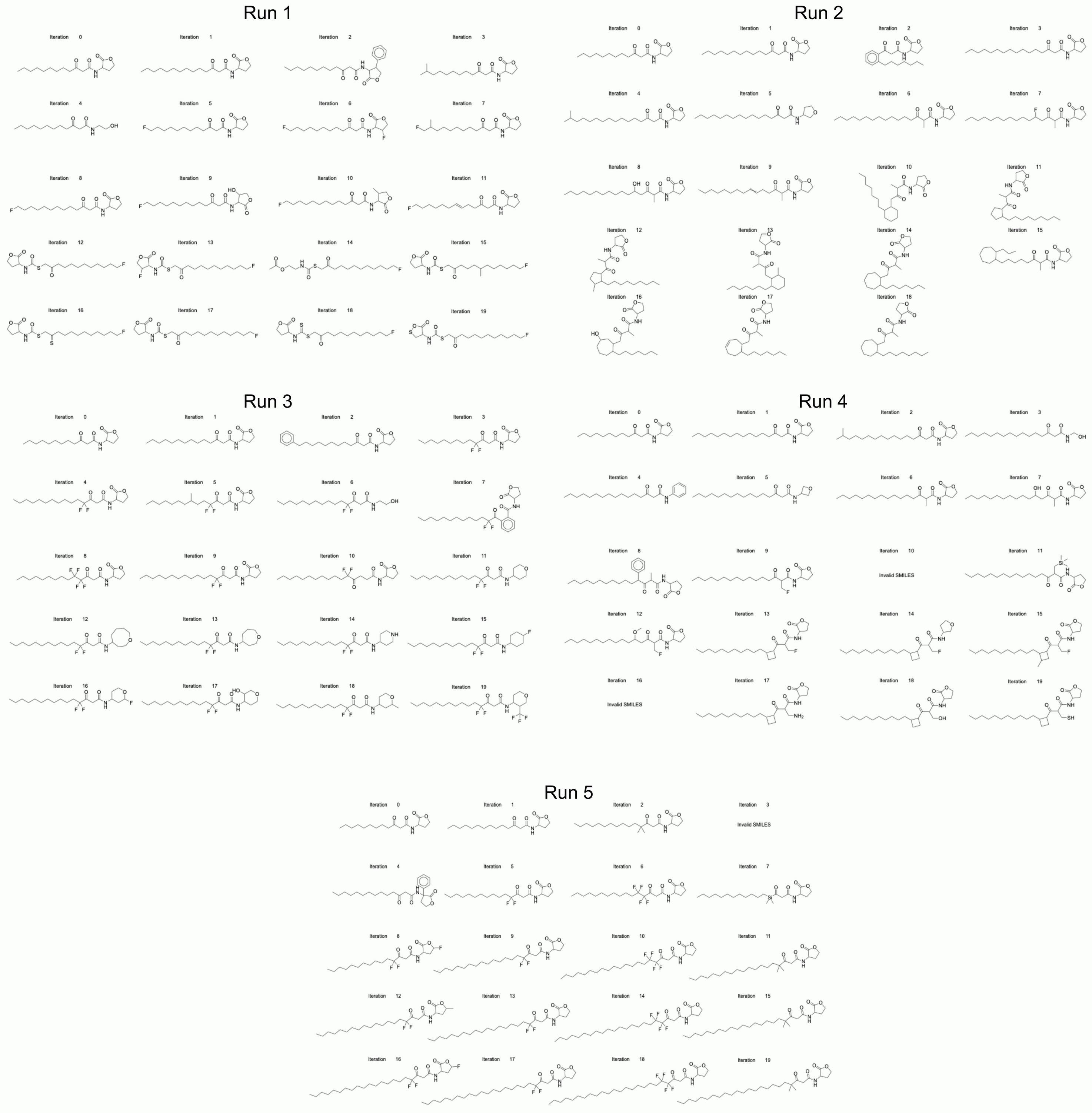}
    \caption{Visualization of the 100 molecules generated in the closed-loop inverse design of surfactants with lower CMC.
    The agent is powered by Claude 3.5 Sonnet.
    No potentially unstable functional groups were found.
    Invalid SMILES generated are marked as invalid.
 \label{fig:cmc_SI}}
\end{figure}

\section{Drug Design and Targeted Docking Inference with GPT-4o\label{sec:si_results_WDR5}}

\begin{figure}[H]
    \centering
    % \captionsetup{justification=centering}
    \includegraphics[width=0.95\textwidth]{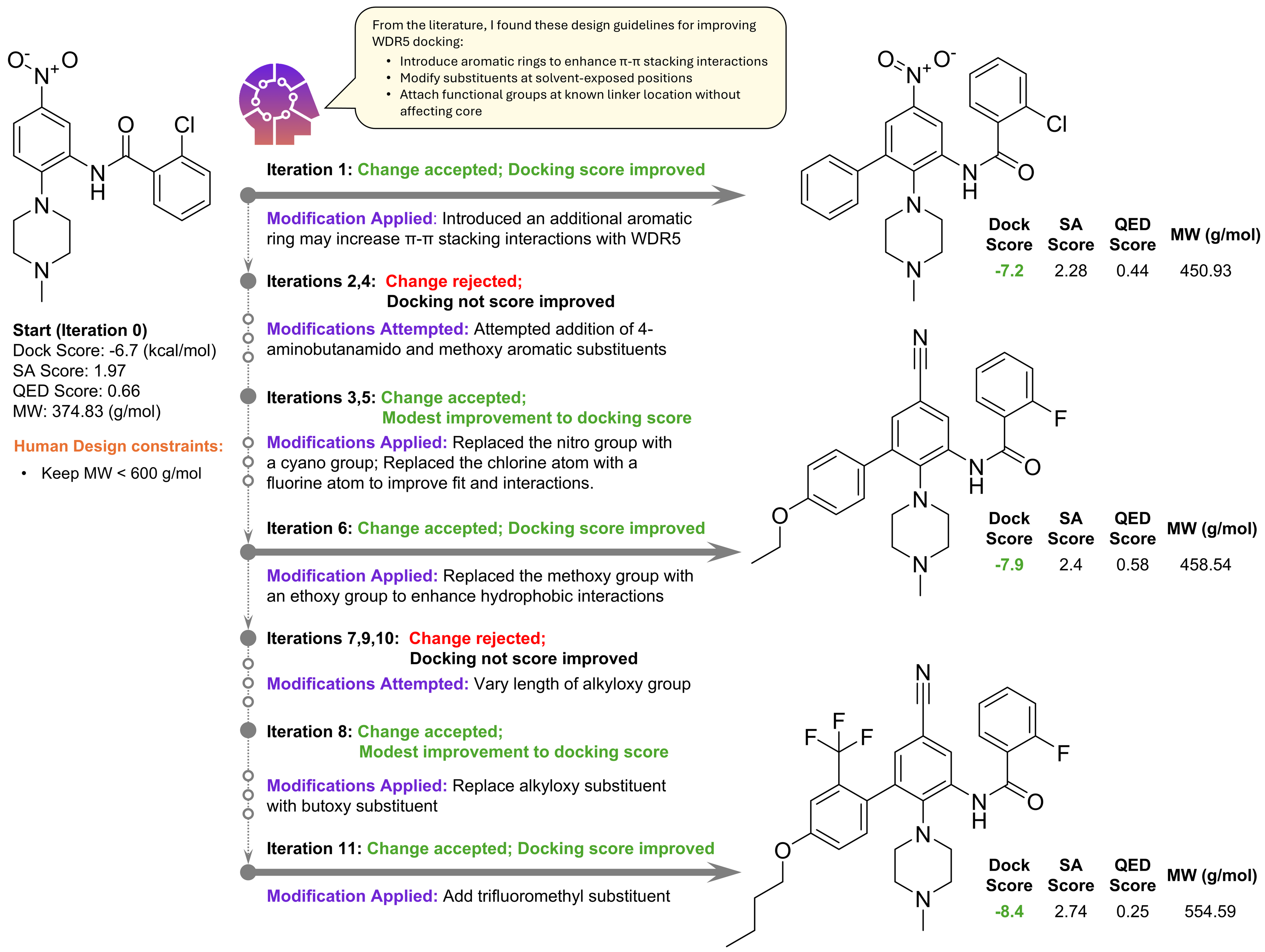}
    \caption{\textbf{dZiner's chain-of-thoughts in the closed-loop inverse design of a drug candidate against WDR5 protein target}.
    The agent is powered by GPT-4o.
    The design guidelines are extracted by the agent from the same references as in Figure 3, and the model is asked to keep the molecular weight lower than 600 (g/mol) in natural language text.
    Docking score is reduced by just over two orders of magnitude via iterative agent-suggested chemical modifications (Dock Score = log(kcal/mol)). The accepted molecule has a Tanimoto similarity score of 0.46 compared to the initial molecule, indicating that substantial changes have been made to the structure in the process of improving binding affinity.
 \label{fig:si_results_docking_gpt4}}
\end{figure}

Similar to Section 2.2, and starting from the same initial HTS hit, we applied dZiner powered by GPT-4o to improve docking against WDR5 (see Figure~\ref{fig:si_results_docking_gpt4}).
In iteration 1, an aromatic ring was added to strengthen hydrophobic interactions, improving the docking score to -7.2. 
Further modifications in iteration 3 included replacing the nitro group with a cyano group, yielding a docking score of -7.4.
In iterations 4-8, functional groups like methoxy, ethoxy, and butoxy were added to enhance hydrophobic interactions, with the best improvement seen in iteration 8, where a butoxy group raised the docking score to -8.2.

Subsequent iterations aimed to fine-tune the structure by replacing the butoxy group with pentoxy and hexoxy groups, though these did not lead to further improvements. 
In iteration 11, a trifluoromethyl group was added to the butoxy-substituted structure, yielding the highest docking score of -8.4. This modification optimized interactions within the binding pocket.
Other attempts, such as adding a trifluoromethoxy group in iteration 18 and a trifluoromethylthio group in iteration 20, showed varying results but did not surpass the best docking score. 
Overall, the case study demonstrated that introducing electron-withdrawing and hydrophobic groups, particularly in iterations 8 and 11, significantly enhanced binding affinity, aligning with the guidelines for targeting WDR5.

\begin{figure}[H]
    \centering
    % \captionsetup{justification=centering}
    \includegraphics[width=\textwidth]{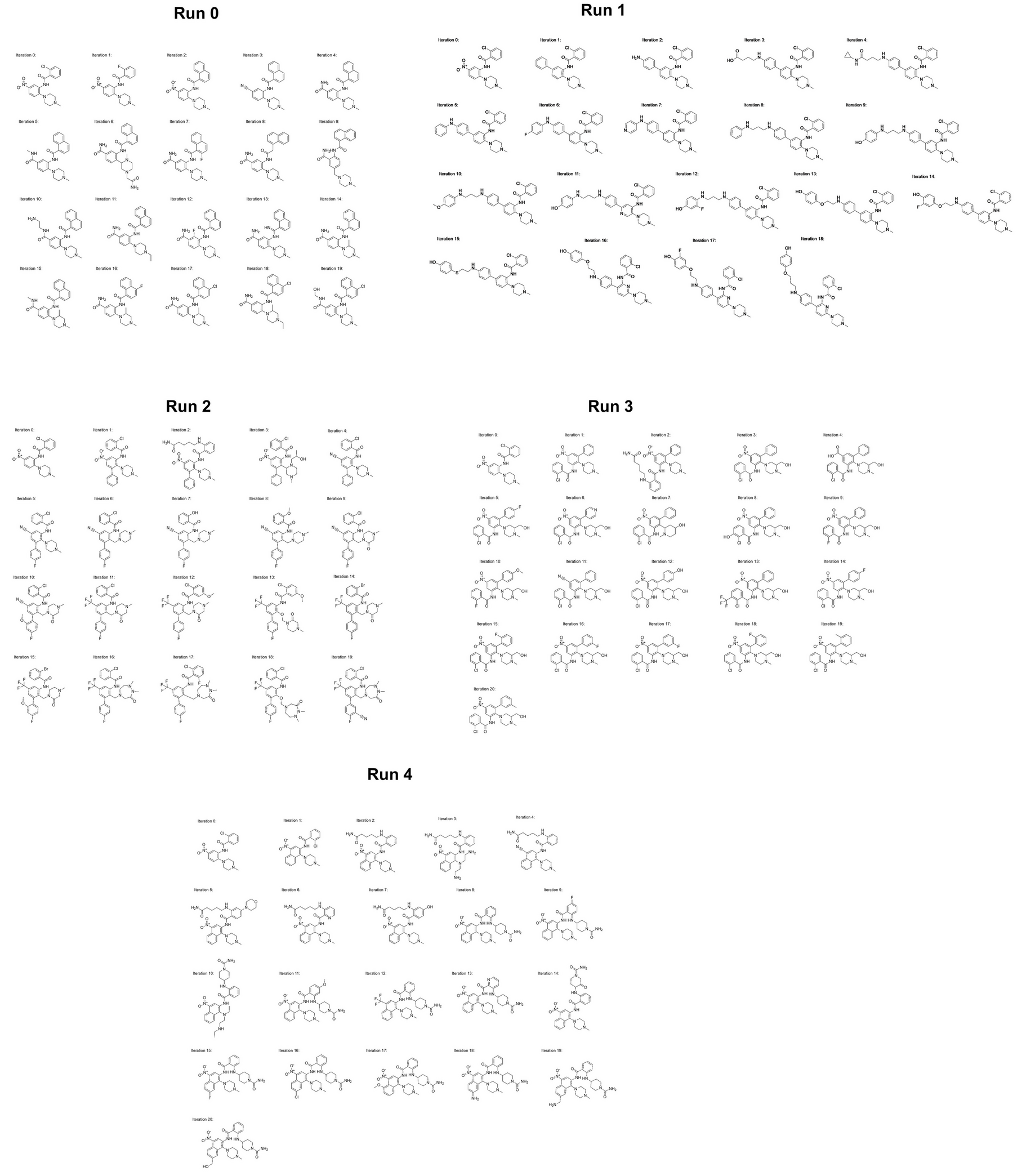}
    \caption{Visualization of the 100 molecules generated in the closed-loop inverse design of drug molecules with high binding affinity against WDR5. The agent is powered by Claude 3.5 Sonnet.
    No molecule was found to be invalid or contain potentially unstable functional groups.
 \label{fig:all_WDR5_SI_with_MW_constraints}}
\end{figure}

\begin{figure}[H]
    \centering
    % \captionsetup{justification=centering}
    \includegraphics[width=\textwidth]{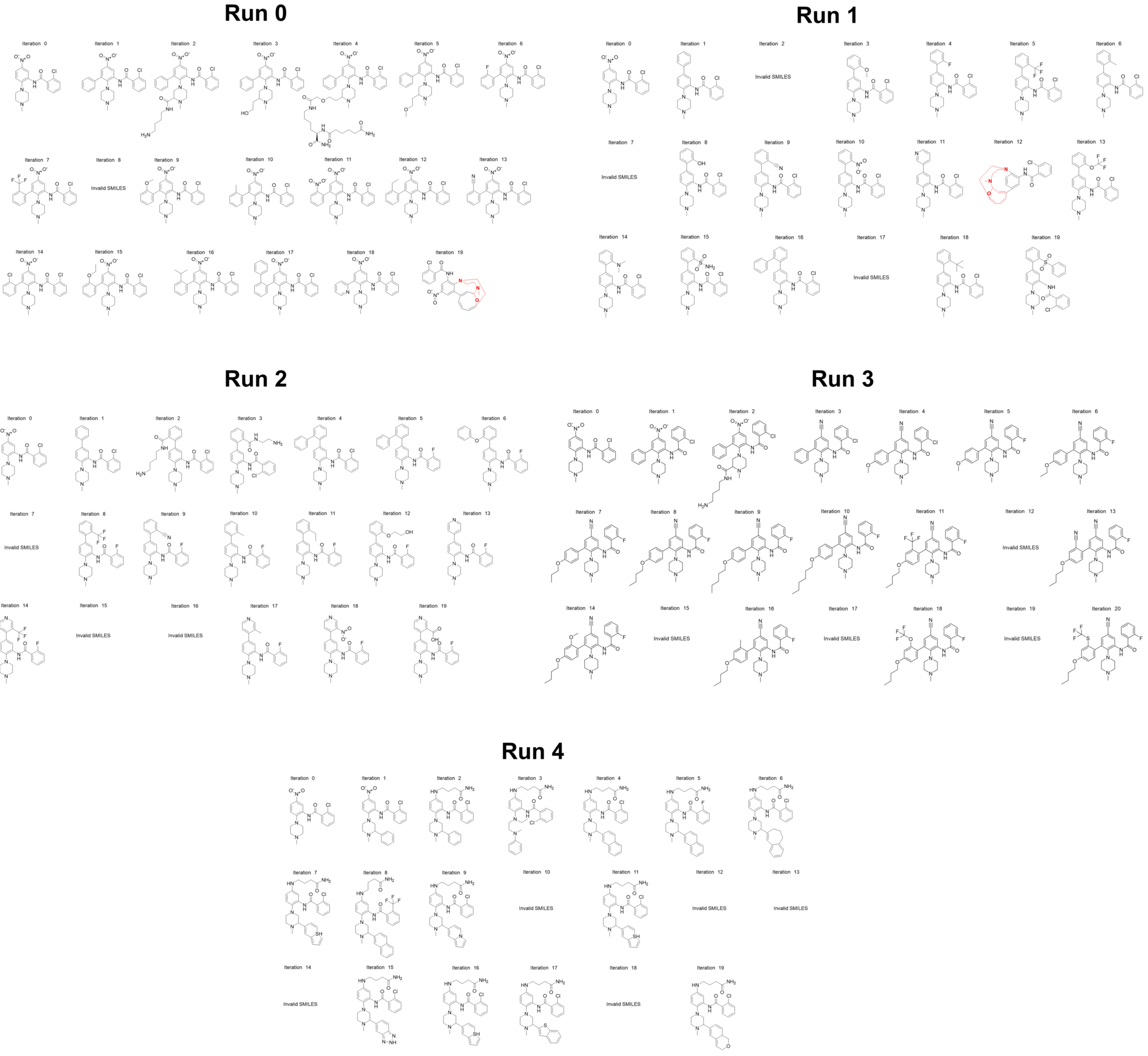}
    \caption{Visualization of the 100 molecules generated in the closed-loop inverse design of drug molecules with high binding affinity against WDR5.
    The agent is powered by GPT-4o.
    Color red indicates a potentially unstable functional group.
    Invalid SMILES generated are marked as invalid.
 \label{fig:all_WDR5_SI_with_MW_constraints}}
\end{figure}

\subsection{Human-in-the-loop design with GPT-4o}
As in closed-loop analysis (section~\ref{sec:si_results_WDR5}), the model was provided an initial structure. Examples of initial modifications include adding hydrophobic groups or optimizing solvent-exposed regions to the same starting molecule. 
After each iteration, the human reviewed the changes, provided specific feedback—such as focusing on different molecular regions or improving solubility—and suggested further modifications like adding functional groups or tweaking $\pi-\pi$ stacking interactions. 
Accepted changes, like adding a phenyl or trifluoromethyl (-CF$_3$) group, increased binding affinity and were incorporated, while less effective modifications were rejected based on the docking score and human input.
Overall, dZiner was cooperative and could be used by medicinal chemists with no prior AI experience. It was able to enable requested most changes, including instructions to revert to prior iterations.  Compared to the closed-loop optimization, human-in-the-loop experiments were able to produce diverse structures with improved WDR5 docking scores.

\begin{figure}[H]
    \centering
    % \captionsetup{justification=centering}
    \includegraphics[width=0.9\textwidth]{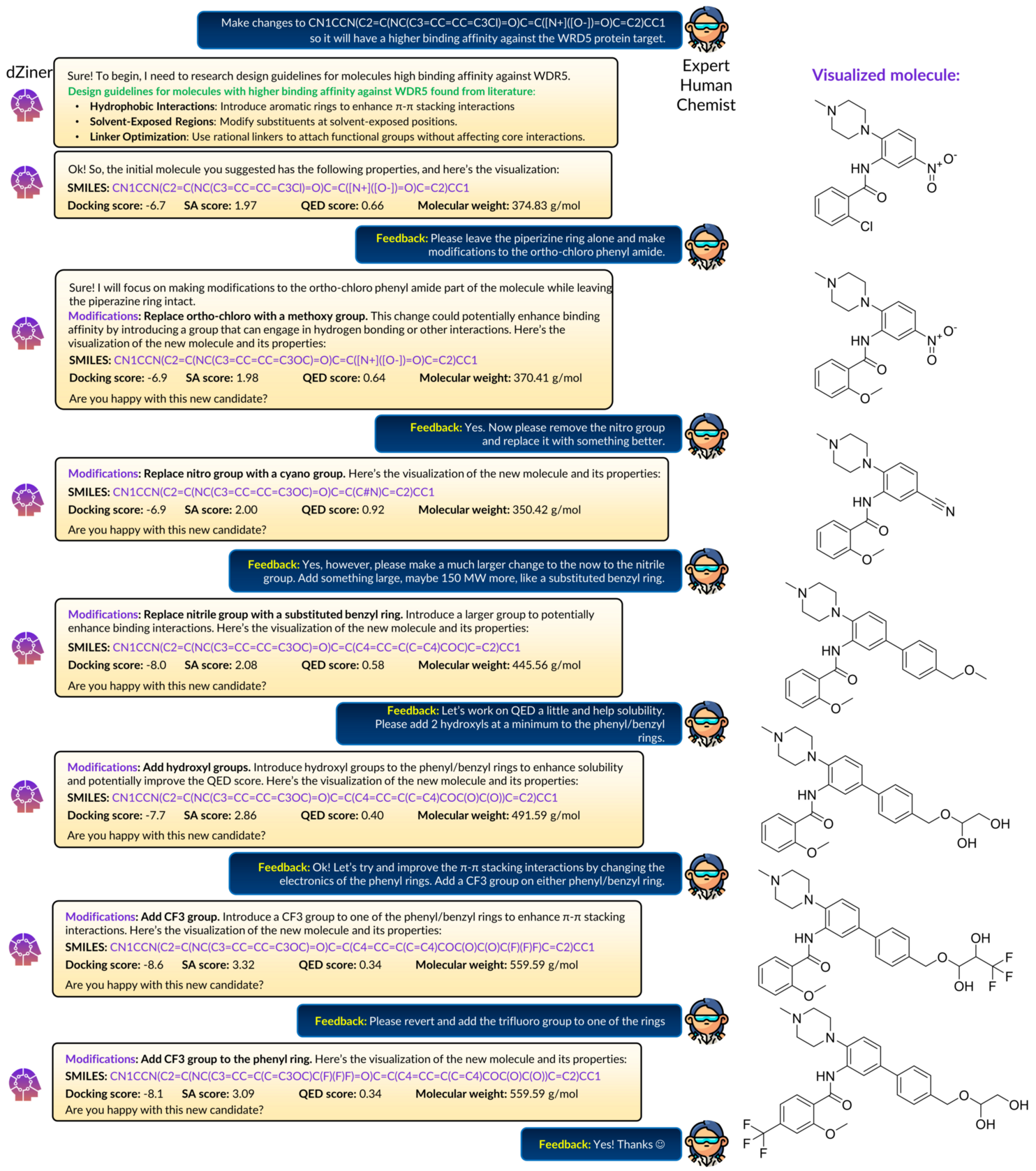}
    \caption{\textbf{Human-in-the-loop inverse design of a drug candidate against WDR5 protein target}.
    Agent is powered by GPT-4o.
    dZiner was found to be highly cooperative, interpretable, and able to enact changes requested with ease in this human-AI collaboration.
 \label{fig:human-in-the-loop}}
\end{figure}

\section{MOF Organic Linker Design and CO$_2$ Adsorption Capacity Inference with GPT-4o\label{sec:SI_results_mof}}
\begin{figure}[H]
    \centering
    % \captionsetup{justification=centering}
    \includegraphics[width=\textwidth]{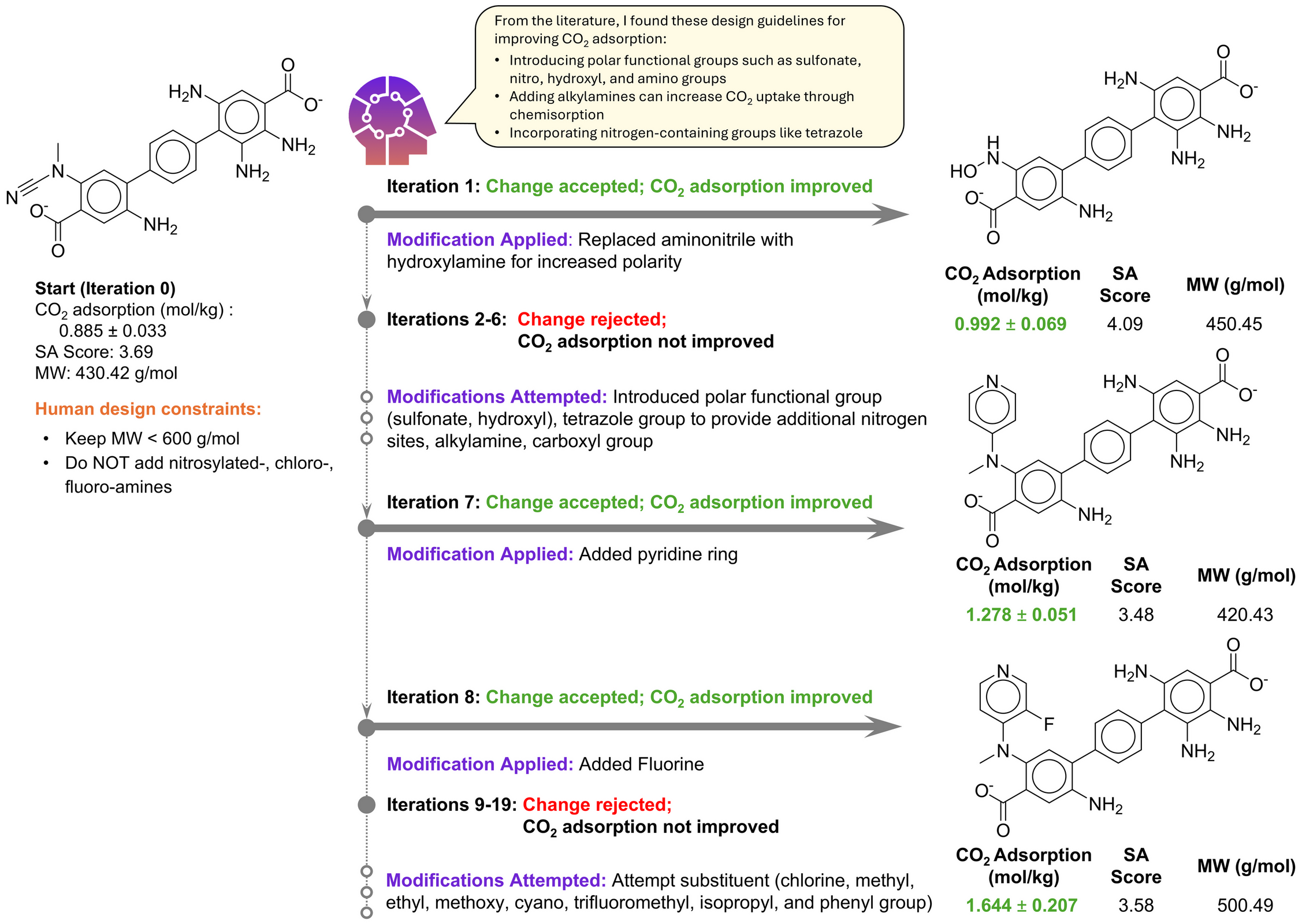}
    \caption{\textbf{dZiner's chain-of-thoughts in the closed-loop inverse design of organic linkers for MOFs with high CO$_2$ adsorption capacity}.
    The agent is powered by GPT-4o.
    Design guidelines were retrieved from scientific literature (same as in Figure 4). 
    The model is asked to keep the molecular weight lower than 600 (g/mol), and \emph{not} to add nitrosylated, chloro-, fluoro- amines to the molecule in natural language text. The accepted molecule bears 63% similarity (Tanimoto) to the starting molecule after 8 iterations, indicating non-trivial modifications have been made to enhance CO2 adsorption.
 \label{fig:CO2_SI_with_all_constraints_gpt4o}}
\end{figure}

Similar to Section 2.3, and starting from the same initial organic linker, we applied dZiner
powered by GPT-4o to enhance CO$_2$ adsorption capacity (see Figure~\ref{fig:CO2_SI_with_all_constraints_gpt4o}). 
In iteration 1, the introduction of hydroxyl (-OH) and amino (-NH$_2$) groups improved the adsorption to 0.992. 
However, in iteration 2, the addition of a sulfonate group resulted in a slight decrease in performance.
Significant improvements were achieved in iteration 7 by incorporating a pyridine ring, which increased nitrogen interactions and boosted CO$_2$ adsorption to 1.278. 
The highest adsorption, 1.644, was observed in iteration 8 when a fluorine atom was introduced, leveraging its high electronegativity to enhance CO$_2$ capture. 
A chlorine atom was also added in iteration 9, resulting in a CO$_2$ adsorption of 1.409. 
Overall, the combination of electronegative atoms and nitrogen-containing functional groups proved most effective in enhancing CO$_2$ adsorption.
Throughout the optimization, the molecular weight increased from 430.424 g/mol (iteration 0) to 516.945 g/mol (iteration 9). The SA score also fluctuated, peaking at 4.595 in iteration 5 after the addition of hydroxyl groups, indicating increased synthetic complexity.

\begin{figure}[H]
    \centering
    % \captionsetup{justification=centering}
    \includegraphics[width=\textwidth]{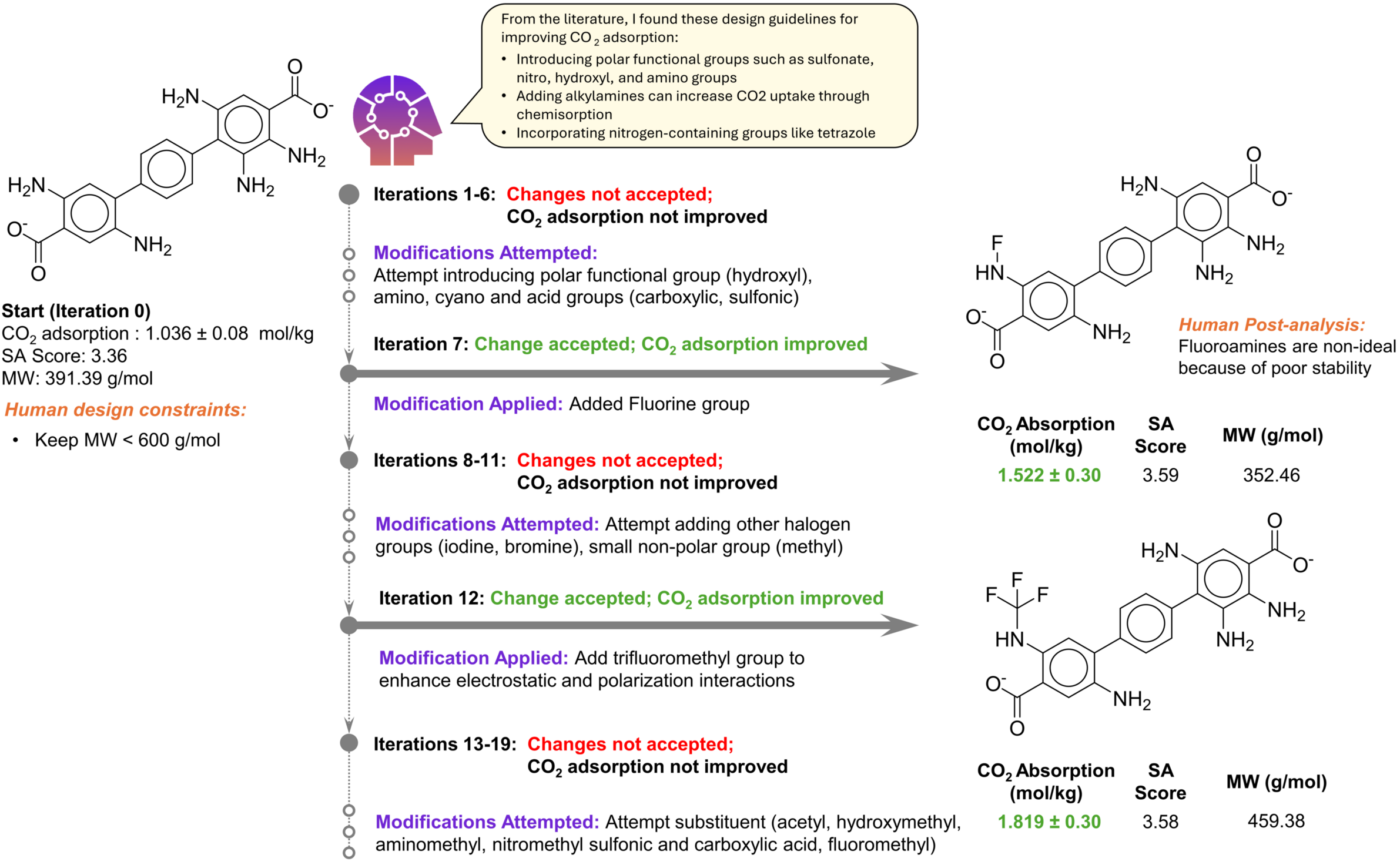}
    \caption{
    \textbf{dZiner's chain-of-thoughts in the closed-loop inverse design of organic linkers for MOFs with high CO$_2$ adsorption capacity}.
    The agent is powered by GPT-4o.
    Design guidelines were retrieved from scientific literature, same as in Figure 4.
    CO$_2$ adsorption capacity is improved by 75\% via iterative agent-suggested chemical
modifications, while following the molecular weight design constraint (MW < 600 g/mol).
 \label{fig:MOFs_SI_gpt_4o_MW_constrain_only}}
\end{figure}

\begin{figure}[H]
    \centering
    % \captionsetup{justification=centering}
    \includegraphics[width=\textwidth]{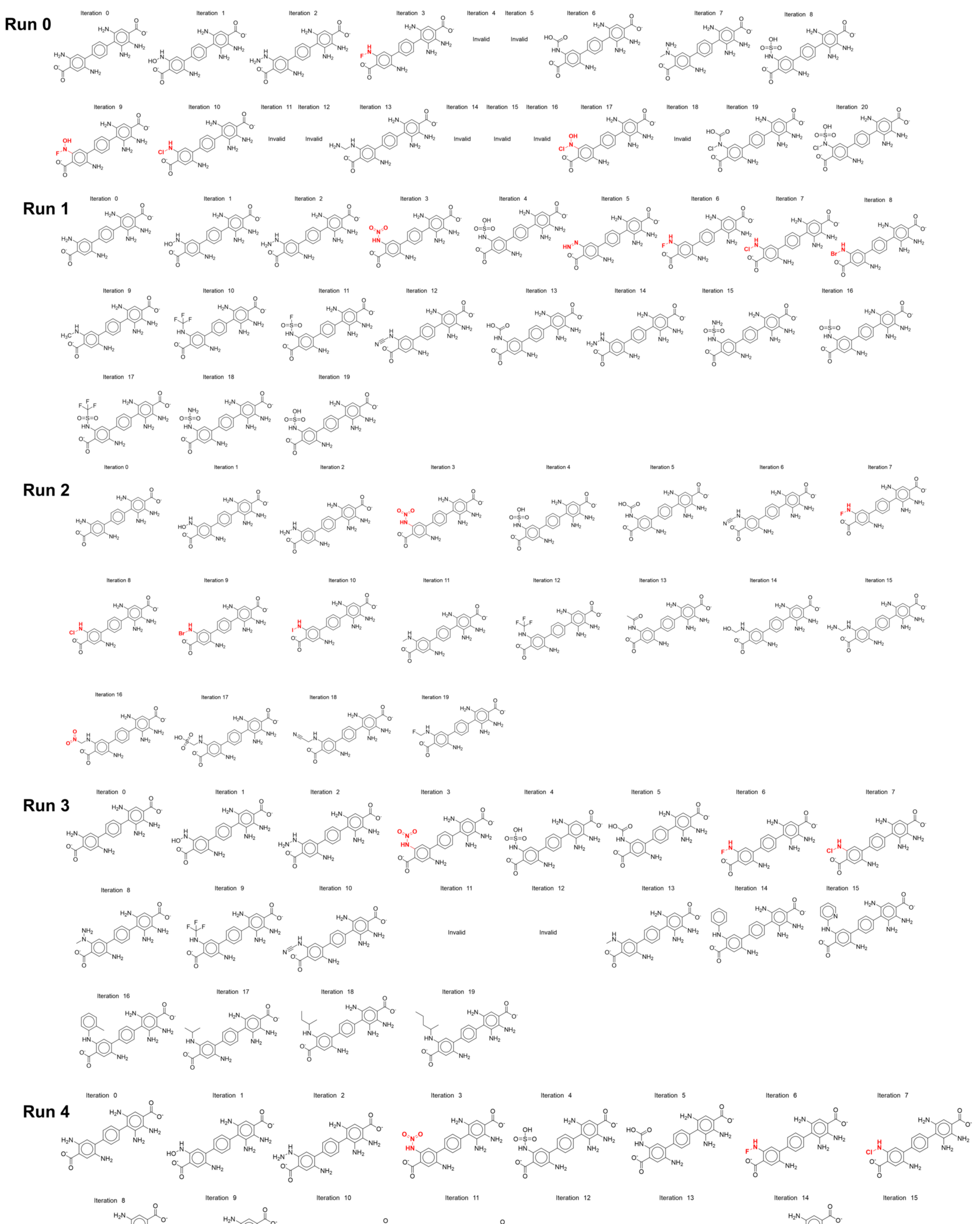}
    \caption{Visualization of the 100 molecules generated in the closed-loop inverse design of organic linkers for MOFs with high CO$_2$ adsorption capacity (molecular weight design constraint only case study).
    The agent is powered by GPT-4o.
    Color red indicates a potentially unstable functional group.
    Invalid SMILES generated are marked as invalid.
 \label{fig:all_CO2_SI_with_MW_constraints}}
\end{figure}

\begin{figure}[H]
    \centering
    % \captionsetup{justification=centering}
    \includegraphics[width=\textwidth]{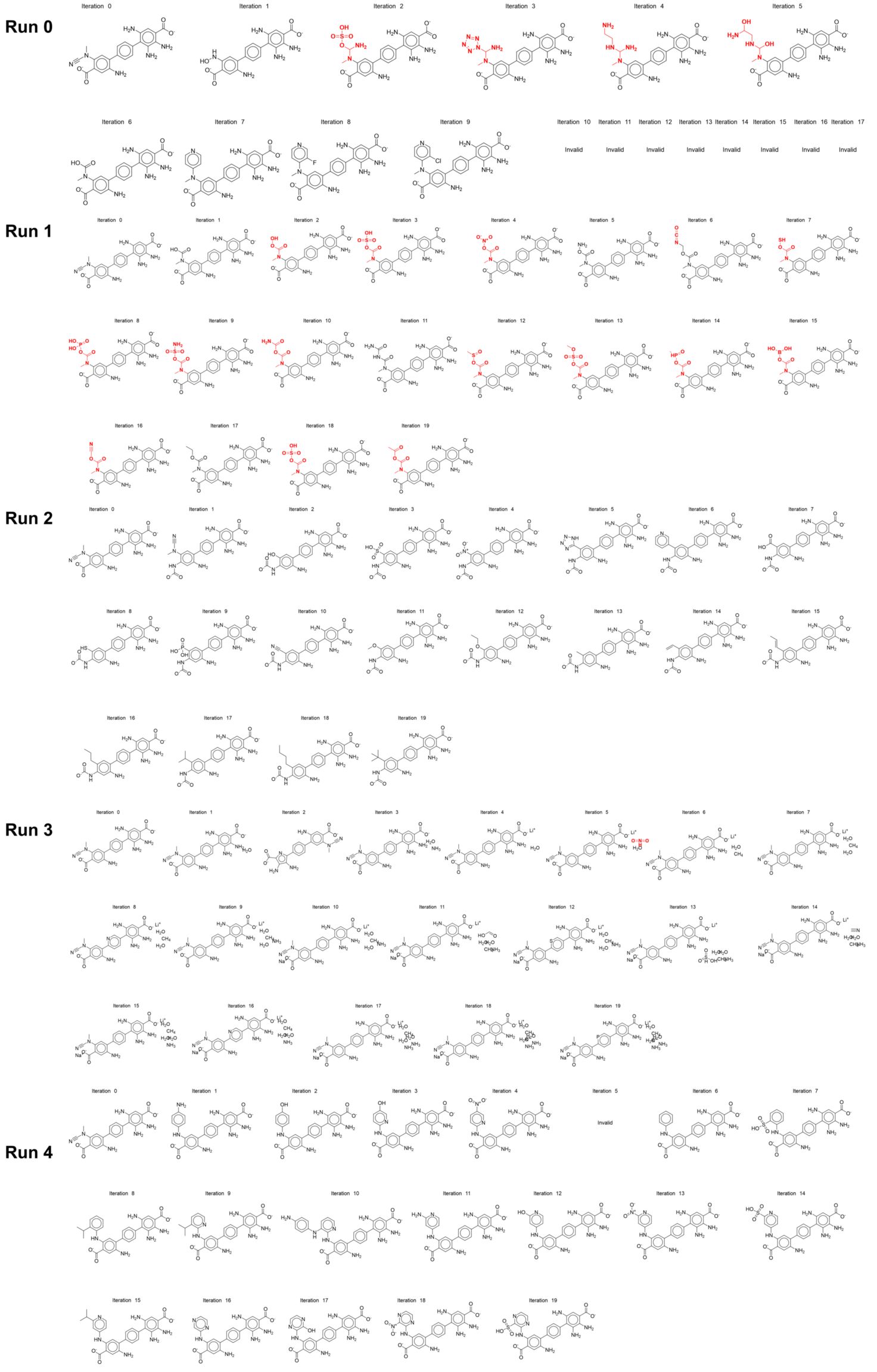}
    \caption{Visualization of the 100 molecules generated in the closed-loop inverse design of organic linkers for MOFs with high CO$_2$ adsorption capacity (molecular weight and functional groups design constraint case study).
    The agent is powered by GPT-4o.
    Color red indicates a potentially unstable functional group.
    No invalid SMILES were generated.
 \label{fig:all_CO2_SI_with_MW_functional_constraints}}
\end{figure}

\begin{figure}[H]
    \centering
    % \captionsetup{justification=centering}
    \includegraphics[width=\textwidth]{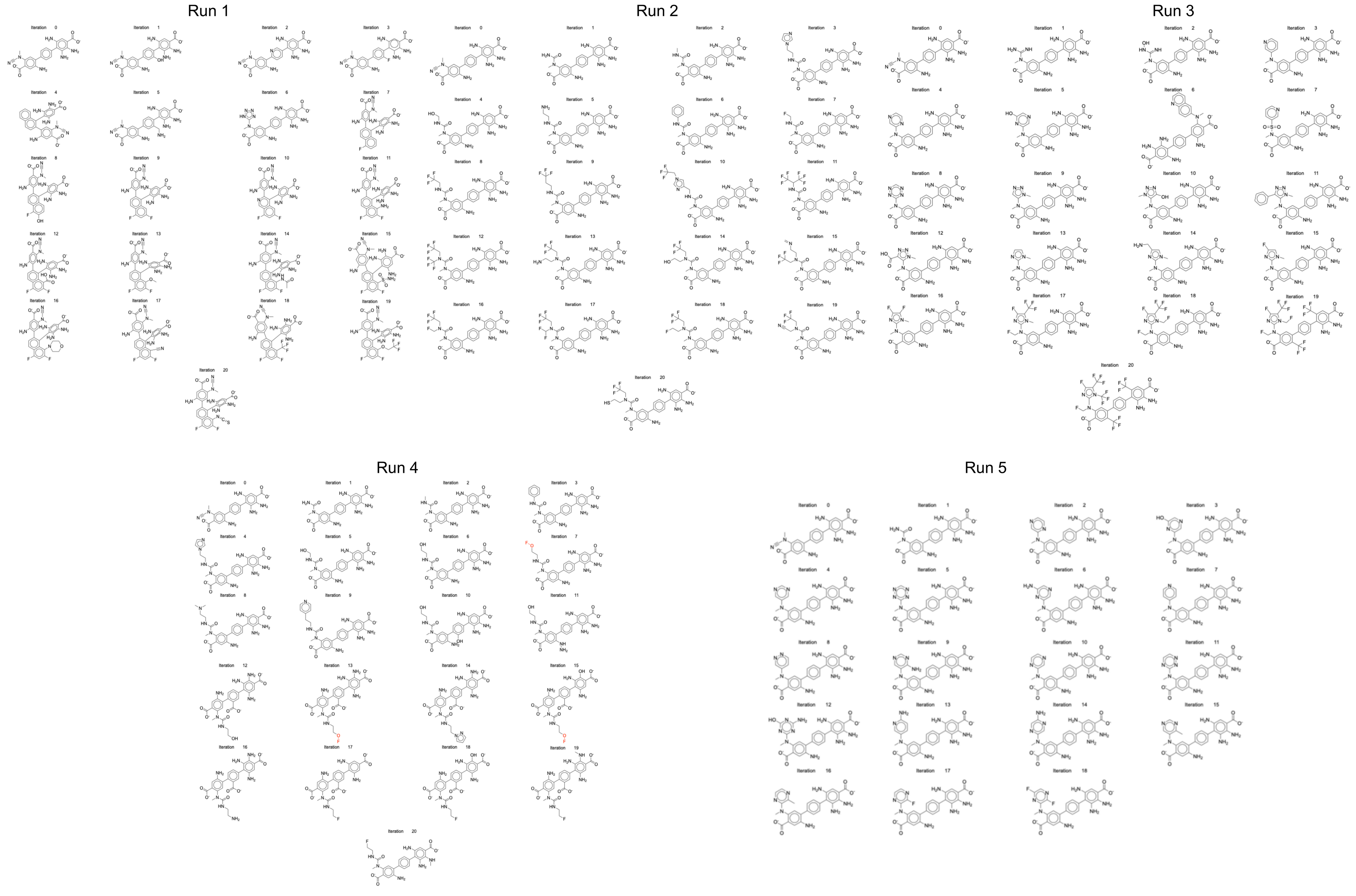}
    \caption{Visualization of the 100 molecules generated in the closed-loop inverse design of organic linkers for MOFs with high CO$_2$ adsorption capacity (molecular weight and functional groups design constraint case study).
    The agent is powered by Claude 3.5 Sonnet.
    Color red indicates a potentially unstable functional group.
    invalid SMILES generated are marked as invalid.
 \label{fig:all_CO2_SI_with_MW_functional_constraints}}
\end{figure}
% \bibliographystyle{unsrtnat}
% \bibliography{bibliography}